\pdfoutput=1

\documentclass[11pt]{article}
\usepackage[final]{acl}

\usepackage{times}
\usepackage{latexsym}

\usepackage[T1]{fontenc}

\usepackage[utf8]{inputenc}

\usepackage{microtype}

\usepackage{inconsolata}

\usepackage{graphicx}

\title{Dialogue Systems Engineering: A Survey and Future Directions}

\author{
  \textbf{Mikio Nakano\textsuperscript{1,2}},
  \textbf{Hironori Takeuchi\textsuperscript{3}},
  \textbf{Sadahiro Yoshikawa\textsuperscript{4}}\\
  \textbf{Yoichi Matsuyama\textsuperscript{4}},
  \textbf{Kazunori Komatani\textsuperscript{1}}
\\
\\
  \textsuperscript{1}SANKEN, University of Osaka, Ibaraki, Osaka, Japan \\
  \textsuperscript{2}C4A Research Institute, Inc., Setagaya, Tokyo, Japan\\
  \textsuperscript{3}Musashi University,  Nerima, Tokyo, Japan\\
  \textsuperscript{4}Equmenopolis, Inc., Shinjuku, Tokyo, Japan\\
 \texttt{mikio.nakano@ei.sanken.osaka-u.ac.jp, h.takeuchi@cc.musashi.ac.jp}\\
 \texttt{yoshikawa@equ.ai, yoichim@equ.ai, komatani@sanken.osaka-u.ac.jp}
  }

\begin{document}
\maketitle
\begin{abstract} 

This paper proposes to refer to the field of software engineering related to the life cycle of dialogue systems as Dialogue Systems Engineering, and surveys this field while also discussing its future directions. With the advancement of large language models, the core technologies underlying dialogue systems have significantly progressed. As a result, dialogue system technology is now expected to be applied to solving various societal issues and in business contexts. To achieve this, it is important to build, operate, and continuously improve dialogue systems correctly and efficiently. Accordingly, in addition to applying existing software engineering knowledge, it is becoming increasingly important to evolve software engineering tailored specifically to dialogue systems. In this paper, we enumerate the knowledge areas of dialogue systems engineering based on those of software engineering, as defined in the Software Engineering Body of Knowledge (SWEBOK) Version 4.0, and survey each area. Based on this survey, we identify unexplored topics in each area and discuss the future direction of dialogue systems engineering.

\end{abstract}

\section{Introduction}

Most recent research on dialogue systems has focused on theoretical and statistical models, machine learning methods and data, user experience, and evaluation of user satisfaction. However, with the advent of large language models (LLMs), building high-performance components has become significantly easier. As a result, the next critical step is to build, operate, and continuously improve dialogue systems in real-world settings. To achieve this, applying existing knowledge of software engineering and evolving software engineering specific to dialogue systems are crucial. Nevertheless, these have not been extensively discussed.

A dialogue system is an interactive system similar to a web service, and also an AI-based system.  However, it differs from typical web services in that it must handle unrestricted natural language input from users. It also differs from other AI systems in that it involves multi-turn interactions. Therefore, it is necessary to advance research on the software engineering of dialogue systems separately from that of general web service systems \cite{web-engineering-review} and AI systems \cite{10.1145/3487043,8804457,8836142}.

In this paper, we refer to the intersection of dialogue system technology and software engineering as {\bf Dialogue Systems Engineering}. 
We first enumerate areas in dialogue systems engineering based on knowledge areas in software engineering, and then present a survey of existing research in each area.
Based on the survey, we identify topics that have not been well explored and discuss future directions for dialogue systems engineering.


Knowledge in dialogue systems engineering has not been sufficiently shared in academia, partly due to the difficulty of quantitative evaluation. However, to promote the development of dialogue system technologies, it is essential to share such knowledge within academia.

This paper deals with all types of dialogue systems, including text-based, speech-based, and multimodal systems, as well as task-oriented and open-domain non-task-oriented systems.

Note that the primary objective of this paper is to identify new research challenges, and thus it does not aim to provide a comprehensive review of all existing work.

\begin{figure}[t]
\begin{center}
\includegraphics[width=0.5\textwidth]{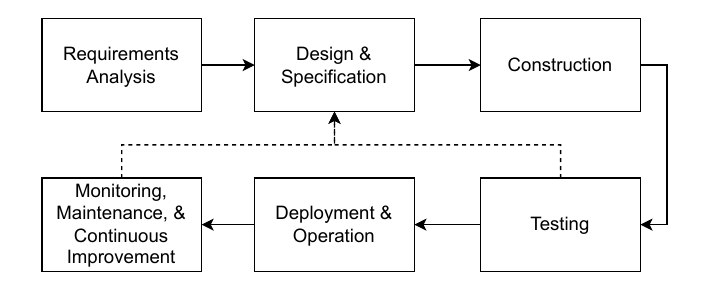}
\end{center}
\caption{A dialogue system life cycle.}
\label{lifecycle}
\end{figure}

\section{Knowledge Areas in Dialogue Systems Engineering}
\label{sec:dse}

Dialogue systems engineering covers all phases of the {\em dialogue system life cycle}. Areas slightly outside the scope of software engineering, such as theoretical or statistical models of dialogue systems, machine learning methods and datasets, and prompt engineering, are excluded. However, techniques and tools used to support the development and improvement of data and models are included within the scope.

We define the dialogue system life cycle as shown in Figure~\ref{lifecycle}, consisting of the following phases: (1) requirements analysis, (2) design and specification, (3) construction, (4) testing, (5) deployment and operation, and (6) monitoring, maintenance, and continuous improvement. Depending on the outcomes of testing and monitoring, the design and specification may be revised.

The dialogue system life cycle has been discussed in previous work, such as \citet{McTear2004} and \citet{dybkjaer2002dialogue}, but Figure~\ref{lifecycle} reflects some modifications. For example, while the model in Figure 6 of \citet{dybkjaer2002dialogue} separates construction and integration, such a separation may no longer be necessary given the advancement of end-to-end models and LLMs. Additionally, the evaluation phase in Figure 6 of \citet{dybkjaer2002dialogue} corresponds to the testing phase in Figure~\ref{lifecycle}.

\begin{table*}[t]
\small
\begin{center} 
\begin{tabular}{rp{4.5cm}p{6cm}p{1.5cm}}
\hline\hline
  & SWEBOK knowledge area  & Dialogue systems engineering knowledge area& Section in this paper \\ \hline\hline
1 & Software requirements & Dialogue system requirements & \ref{requirements}   \\ \hline
2 & Software architecture & Dialogue system architecture & \ref{architecture} \\ \hline
3 & Software design  & Software design of dialogue systems & \ref{design} \\\hline
4 & Software construction  & Dialogue system construction & \ref{construction} \\ \hline
5 & Software testing & Dialogue system testing &  \ref{test} \\ \hline
6 & Software engineering operations & Deployment and operation of dialogue systems & \ref{operations} 
\\\hline
7 & Software maintenance & Monitoring, maintenance, and continuous improvement of dialogue systems & \ref{maintenance} 
\\\hline
8 & Software configuration management &  & \\\hline
9 & Software engineering management & & \\\hline
10 & Software engineering process  & Dialogue system life cycle & \ref{sec:dse}  \\\hline
11 & Software engineering models and methods \\\hline
12 & Software quality & Quality of dialogue systems & \ref{quality} \\\hline
13 & Software security & Security, safety, and privacy protection of dialogue systems & \ref{security} \\\hline
14 & Software engineering professional practice & Dialogue systems engineering professional practice & \ref{pracice} \\   \hline
15 & Software engineering economics & Dialogue system economics & \ref{economics} \\\hline
16 & Computing foundations      &     \\ \hline
17 & Mathematical foundations   &        \\ \hline
18 & Engineering foundations    &       \\ \hline\hline
\end{tabular}
\end{center}
\caption{Knowledge areas of Dialogue Systems Engineering. The leftmost column represents the SWEBOK chapter number. If the ``Dialogue systems engineering knowledge area'' column is empty, it means there are no dialogue system-specific topics.}
\label{dse-ka}
\end{table*}

We based our definition of knowledge areas in dialogue systems engineering on the knowledge areas defined in the Guide to the Software Engineering Body of Knowledge ver. 4.0 (SWEBOK hereafter) by the IEEE Computer Society \cite{swebok4}. SWEBOK comprehensively covers the domains of software engineering knowledge. Table~\ref{dse-ka} shows how each SWEBOK knowledge area maps to areas in dialogue systems engineering. If the ``Area in Dialogue Systems Engineering'' column is left blank, it indicates that we found no dialogue system-specific knowledge area corresponding to that SWEBOK category. However, the absence of a dialogue system-specific knowledge area does not imply that software engineering knowledge is irrelevant. Rather, general software engineering practices remain directly applicable to dialogue system development and operation.
Note that the dialogue system life cycle has already been discussed in this section and is therefore not included in the survey in the next chapter.

As SWEBOK itself notes interdependencies among its knowledge areas, it is important to recognize that the knowledge areas of dialogue systems engineering in Table~\ref{dse-ka} are also interconnected. For example, to ensure privacy in a dialogue system, one might adopt a client-server architecture that avoids transmitting personal data to the server.

\section{A Survey on Dialogue Systems Engineering} \label{survey}

\subsection{Survey Method}

This survey was conducted based on the authors' domain knowledge as well as keyword searches using the names of knowledge areas and related terms on platforms such as Google Scholar\footnote{\url{https://scholar.google.com/}} and DBLP\footnote{\url{https://dblp.org/}}. In addition, we used tools like OpenAI's ChatGPT Search\footnote{\url{https://openai.com/index/introducing-chatgpt-search/}} and Deep Research\footnote{\url{https://openai.com/index/introducing-deep-research/}}, although the results from these tools were reviewed and summarized rather than used verbatim in this paper.

We did not conduct a {\em systematic literature review} \cite{CARRERARIVERA2022101895} because the knowledge areas in dialogue systems engineering are highly diverse and span multiple academic disciplines. As a result, terminology varies widely, and it is difficult to comprehensively collect relevant papers using keywords alone.

\subsection{Dialogue System Requirements} \label{requirements}

Dialogue system requirements define what the dialogue system should do, what services it provides, and what constraints it must satisfy.\citet{McTear2004} discusses requirements analysis by dividing it into two categories: use case analysis and spoken language requirements. \citet{Sonntag:2010} conducted requirements analysis for industrial applications based on usability analysis and use case analysis. Additionally, \citet{ATWEL2000} emphasize the need for involving domain experts in the specification of dialogue system requirements.

Requirements specifications are closely tied to the evaluation of dialogue systems. Common evaluation metrics include {\em user satisfaction} \cite{walker-etal-1997-paradise,ultes:dd21} and {\em user experience} \cite{clark:chi19,Johnston2023}. However, relying solely on user satisfaction and experience overlooks aspects related to system development and operation. To address this, \citet{nakano:iwsds24} propose evaluating from the system owner's perspective. Furthermore, \citet{nakano:iwsds25} introduce an approach that applies business-IT alignment models to comprehensively list the value, cost, and risks of dialogue systems from multiple viewpoints.

\paragraph{Unexplored Issues}

While it is essential to define requirements specifications based on such multifaceted evaluations, concrete methodologies for creating requirements specifications for dialogue systems have not been extensively discussed.

\begin{figure}[t]
\begin{center}
\includegraphics[width=0.5\textwidth]{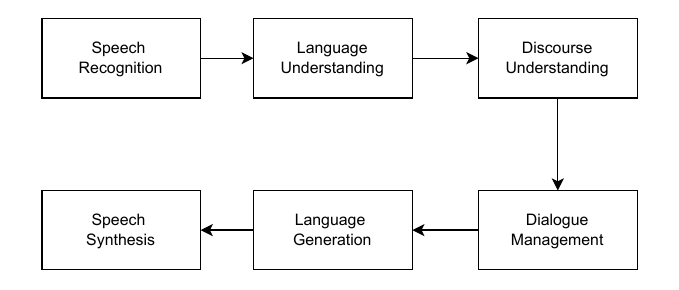}
\end{center}
\caption{A pipeline architecture for dialogue systems.}
\label{pipeline}
\end{figure}

\subsection{Dialogue System Architecture} \label{architecture}

Software architecture can be broadly defined as the roles of modules within software and the communication between them. In dialogue systems, architecture plays a critical role due to the need to integrate a wide range of technologies.

A widely used architecture for dialogue systems consists of sequential components such as speech recognition (in spoken dialogue systems), language understanding, discourse understanding (or dialogue state tracking for statistical systems), dialogue management (or action selection), language generation, and speech synthesis (in spoken dialogue systems) \cite{ALLEN01011995,DBLP:conf/aaai/FergusonA98,DBLP:conf/interspeech/SeneffHLPSZ98}. This is commonly referred to as a {\em pipeline architecture} (Figure~\ref{pipeline}). Variants of this architecture are also used in modern toolkits such as ConvLab \cite{zhu-etal-2023-convlab} and Rasa Open Source \cite{bocklisch2017rasaopensourcelanguage}.

Based on this architecture, many variations have been proposed, including architectures designed for domain and language portability, distributed architectures for multi-domain dialogue systems, architectures for speech and multimodal systems, and client-server architectures.
In recent years, architectures consisting solely of modules that generate utterances end-to-end, such as those employing large language models (LLMs), have also been used, differing from the traditional pipeline architecture.

Appendix~\ref{appendix:architecture} lists various architectures for dialogue systems.

\paragraph{Unexplored Issues}

In dialogue systems research, architectures have often been evaluated solely based on metrics such as the usability of the application systems, and they are rarely assessed from the perspective of the entire dialogue system life cycle.
In software engineering, for example, {\em cohesion} (the degree of functional purity within a module) and {\em coupling} (the degree of interdependence between modules) are commonly used evaluation criteria \cite{structured-design}. High cohesion and low coupling are desirable. The basic pipeline architecture satisfies these criteria. This is because each module operates with a high degree of independence and minimal inter-module communication. Consequently, individual modules can be easily replaced. In fact, \citet{takanobu-etal-2020-goal} demonstrate that various modules can be substituted and system-wide performance can still be effectively measured.
In this way, it is beneficial to discuss dialogue system architectures from the perspective of software engineering, and this would be the case for architectures for dialogue systems that leverage LLMs.

\subsection{Software Design of Dialogue Systems} \label{design}

Software design of a dialogue system means the design of individual modules within it.
Several software design methods proposed in software engineering have been applied to dialogue system development. For example, \citet{ONEILL_McTEAR_2000} introduced object-oriented design in distributed multi-domain dialogue systems. This approach enables the separation of domain-independent and domain-specific processes, facilitating more efficient development.

In spoken and multimodal dialogue systems, user speech and multimodal events are often input asynchronously, requiring internal module communication also to be asynchronous. To support this, event-driven design approaches have been adopted \cite{wang1998event,hartholt:iva13,meng:isis}.

\paragraph{Unexplored Issues}

In addition to the methods mentioned above, software design approaches such as domain-driven design and aspect-oriented design have also been proposed. However, there has been little research on their application to dialogue systems.

\subsection{Dialogue System Construction} \label{construction}

Building dialogue systems requires integrating a variety of technologies, and the construction process can be costly. Therefore, research has been conducted to facilitate and streamline this process.

\paragraph{Dialogue system development tools}
One line of research focuses on development tools for dialogue systems. Numerous tools, both for research and commercial use, have been developed.

For example, the CSLU Toolkit \cite{cslu-tookit,sutton1998universal} was an early tool for building task-oriented multimodal dialogue systems. It included components such as speech recognition, speech synthesis, and facial display, and enabled developers to design dialogue management through GUI-based state transitions. For non-task-oriented text-based systems, AIML (Artificial Intelligence Markup Language) interpreters \cite{wallace2009anatomy} have been used. Additionally, tools have been developed for various types of dialogue systems. Those include tools for developing dialogue systems that employ LLMs (see Appendix~\ref{appendix:tools}).

Using these tools, dialogue systems can be developed with no-code or low-code approaches.

\paragraph{Development methodology}
Tools such as Xatkit \cite{xatkit} and Jarvis \cite{Jarvis} apply {\em model-driven development} (MDD) using a {\em domain specification language}. \citet{DBLP:conf/modelsward/VahdatiR24} propose a new development approach that combines model-driven development with {\em microservice architecture} to address the complexity of chatbot development, including the variety of technologies and tool dependencies. \citet{araki-2012-rapid} also presents a model-driven development method that automatically constructs dialogue systems from semantic resources. Development using VoiceXML\footnote{\url{https://www.w3.org/TR/voicexml21/}} can also be considered a form of model-driven development. \citet{lintest} have developed a testing tool to support {\em test-driven development}.

\paragraph{Data collection}

Since statistical methods have become widely used in dialogue system modules, collecting human-system dialogue data has become essential for training models. However, it is difficult to observe how users will interact with a dialogue system before it is implemented. To address this, the {\em Wizard-of-Oz} (WoZ) method has been widely adopted. In a WoZ experiment, the wizard needs to quickly understand user utterances and produce appropriate system responses, which is challenging. Therefore, tools that support WoZ experiments are highly useful. \citet{10.1093/iwc/iwu016} list various WoZ tools.

{\em Crowdsourcing} is also an important approach to data collection. \citet{huynh-etal-2022-dialcrowd} and \citet{manuvinakurike-etal-2015-reducing} have developed tools for collecting dialogue data via crowdsourcing. \citet{kang-etal-2018-data} compare datasets collected using different crowdsourcing strategies.

\paragraph{Unexplored Issues}

Not many studies on methods for making dialogue system construction more efficient have been published, partly due to the difficulty of quantitative evaluation. However, sharing insights gained from applying software engineering methodologies, such as model-driven development and test-driven development, to the construction of various types of dialogue systems is highly valuable. In particular, it is important to discuss how to improve the efficiency of building dialogue systems based on large language models (LLMs), which have become prominent in recent years. Since there are still few development tools available for dialogue systems that utilize LLMs, further research in this area is necessary.

\subsection{Dialogue System Testing} \label{test}

Software testing includes two types: {\em static testing}, which involves reviewing specifications, design documents, and source code, and {\em dynamic testing}, which involves executing the software. In this section, we focus on dynamic testing for dialogue systems, particularly {\em integration testing}, which is a major challenge.

Testing dialogue systems differs significantly from testing other types of software due to two key characteristics: the input is unrestricted, and the interaction involves multiple conversational turns. Consequently, {\em online testing}, where human users interact with the system repeatedly, is often conducted.

One approach to online testing is recruiting test users via crowdsourcing. \citet{li-etal-2021-legoeval} developed a toolkit for testing and evaluating text-based dialogue systems using crowdsourcing. In contrast to text-based dialogue systems, spoken and multimodal dialogue systems have traditionally required test users to visit a lab. However, \citet{jurcicek2011real} developed a platform for conducting spoken dialogue system tests via crowdsourcing. \citet{aicher:iwsds23} and \citet{ramanarayanan2017crowdsourcing} discuss various issues encountered when testing spoken and multimodal dialogue systems through crowdsourcing. \citet{manuvinakurike-etal-2015-reducing} analyze the cost of evaluation via crowdsourcing.

It is known that users behave differently depending on whether they are participating in a test or using the system with a real purpose \cite{ai-etal-2007-comparing}. To facilitate evaluation with real users, Let's Go Lab \cite{lets-go-lab} was developed as a platform that allows researchers to deploy their dialogue systems for real-world use. In this system, users call a bus schedule information service, allowing researchers to evaluate and compare systems in a real-world setting. Similarly, DialPort \cite{zhao-etal-2016-dialport,lee-etal-2017-dialport,mehri-eskenazi-2020-unsupervised,huynh-etal-2022-dialport} is a portal that aggregates spoken dialogue systems developed by research institutions, enabling the collection of dialogue data from real users and facilitating comprehensive testing of diverse systems.

Because user-based testing is expensive and time-consuming, various testing tools have been proposed. LINTest \cite{lintest} supports {\em test-driven development} by verifying system behavior against a dialogue corpus. \citet{atefi2019automatedtestingframeworkconversational} also proposes an automated testing framework. Additionally, \citet{guo2024mortarmetamorphicmultiturntesting} apply {\em metamorphic testing} \cite{chen2020metamorphictestingnewapproach}, and \citet{mutation-test} introduce a {\em mutation testing} approach \cite{mutation-testing-survey}. Various other tools have also been developed; \citet{li-review} survey both commercial and academic testing tools.

To detect unexpected problems that occur when predefined test cases are absent, user simulators can be used \cite{Schatzmann:survey,ai-litman-2009-setting}, although user simulators are primarily used for training and evaluating dialogue strategies \cite{pietquin2013survey}. With recent advances in LLMs, several LLM-based user simulators have been proposed \cite{ALGHERAIRY2025101697, luo-etal-2024-duetsim}, and they are also useful for testing. \citet{nakano-sigdial25} propose a method for automatically generating personas for LLM-based user simulators for testing interview dialogue systems.
To identify issues in dialogues generated through user simulation, it is necessary to examine the content of the dialogue itself, except for obvious issues such as system crashes. While manual inspection by humans is one possible method, it is labor-intensive. Therefore, automatic methods for detecting such issues have been proposed \cite{Higashinaka2021,finch-etal-2023-leveraging}.

\paragraph{Unexplored Issues}

Testing is important in the development of dialogue systems, but research on testing tools is still limited. In particular, there is a need for research on testing tools for spoken and multimodal dialogue systems, as well as non-task-oriented dialogue systems. Moreover, how to evaluate the effectiveness of such testing tools remains an open issue.

\subsection{Deployment and Operations of Dialogue Systems} \label{operations}

As with other web services, deploying dialogue systems involves a range of issues such as security, scalability, and cost. In particular, spoken and multimodal dialogue systems require more specialized expertise than web applications or text-based dialogue systems, due to the increased complexity of communication channels with users.

Commercial dialogue system development platforms provide features that simplify the deployment of text and spoken dialogue systems. However, there has been little research discussing deployment from the perspective of software engineering.

\citet{leuski-artstein-2017-lessons} describe the issues encountered and solutions implemented when deploying and operating the New Dimensions in Testimony system \cite{traum:ndt} in a museum setting. The initial system was a combination of research tools, but because technical expertise was not readily available in the museum environment, the system was restructured into a single integrated application to facilitate installation and operation. The authors also discuss efforts such as logging conversations to a server. \citet{afzal-etal-2019-development} use a microservice architecture and orchestration to simplify the deployment of an intelligent tutoring system. Similarly, \citet{ROCA2020103305} adopt a microservice architecture to ensure scalability for a chatbot designed to support chronic patients.

\paragraph{Unexplored Issues}

While deployment and operational knowledge for dialogue systems is not frequently shared, issues such as security and operational costs are non-trivial. In particular, the management of incidents unique to dialogue systems, such as erroneous responses due to system faults, is essential. Sharing practical knowledge and experiences in this area is highly desirable.

\subsection{Monitoring, Maintenance, and Continuous Improvement of Dialogue Systems} \label{maintenance}

This section discusses methods for improving dialogue systems after deployment. Because dialogue systems must handle a wide variety of user utterances, and it is impossible to predict all such inputs before deployment, it is necessary to collect interaction logs post-deployment, identify issues, and revise the system accordingly, sometimes even modifying the specifications.

Several studies have explored the automation of issue detection in deployed dialogue systems \cite{jacovi2020improving, Andrist17, pang-etal-2024-leveraging}.

In software engineering, the concept of DevOps has been proposed to integrate system operation and development into a continuous process \cite{devops:csur}. MLOps \cite{10.5555/2969442.2969519} extends this concept to machine learning systems. Building on these ideas, \citet{yoshikawa:dialops} propose DialOps as a framework for the continuous development and operational management of dialogue systems. They highlight several unique aspects of dialogue systems compared to general MLOps, such as the impact of non-model factors like network latency and user preferences, the real-time nature of interactions, and the diverse temporal dimensions involved in evaluation.

\paragraph{Unexplored Issues}

With the advancement of LLMs, it is desirable to see an increase in research on monitoring, maintenance, and continuous improvement in LLM-based dialogue systems.
In particular, hallucination is a significant issue in the case of LLMs. Incorporating automatic hallucination detection methods \cite{honovich-etal-2022-true-evaluating,niu-etal-2024-ragtruth,song-etal-2024-rag,farquhar2024detecting} into the monitoring of deployed dialogue systems is one of the important topics.

\subsection{Quality of Dialogue Systems} \label{quality}

Software quality in dialogue systems refers to the evaluation criteria that should be specified in the system requirements and represents what should be tested using the methods discussed in Section~\ref{test}.

Section 6.2 of the systematic survey by \citet{software-based-ds} reviews research on the quality of dialogue systems. It classifies the evaluation criteria used in those studies based on the eight software quality characteristics defined in ISO/IEC 25010:2011\footnote{This version has been withdrawn. In the published version ISO/IEC 25010:2023 \cite{ISO/IEC25010:2023}, {\em usability} and {\em portability} have been replaced with {\em interaction capability} and {\em flexibility}, respectively.} \cite{ISO/IEC25010:2011}: {\em functional suitability}, {\em performance efficiency}, {\em compatibility}, {\em usability}, {\em reliability}, {\em security}, {\em maintainability}, and {\em portability}. Among these, the survey categorizes prior studies into four quality characteristics: functional suitability, performance efficiency, usability, and security.

The first three correspond to what has been traditionally referred to in dialogue systems research as user satisfaction \cite{walker-etal-1997-paradise,ultes:dd21} and user experience \cite{clark:chi19,Johnston2023}. Security will be discussed in more detail in Section~\ref{security}.

\paragraph{Unexplored Issues}

The survey by \citet{software-based-ds} only covers four of the eight ISO/IEC 25010:2011 quality characteristics, and does not mention studies addressing compatibility, reliability, maintainability, or portability. In the actual design and construction of dialogue systems, it is important to include these quality characteristics in the system specification. Further research is needed to explore how these aspects should be evaluated.

\subsection{Security, Safety, and Privacy Protection of Dialogue Systems} \label{security}

Here we deal with conditions (as quality attributes) and methods to ensure the security, safety, and privacy of dialogue systems. Various studies have addressed them.

Regarding security, \citet{chatbot-security-privacy} summarize various attack methods and their countermeasures, organizing them by client, response generation module, communication between the client and the response generation module, and database. Recently, {\em prompt injection attacks} targeting LLMs have also become a critical issue \cite{liu2024promptinjectionattackllmintegrated}.
Guardrails such as Nemo Guardrails \cite{rebedea-etal-2023-nemo} have been developed to protect from such attacks.

Safety in dialogue systems includes aspects such as avoiding the provision of harmful or incorrect information (hallucinations). Various methods have been proposed to prevent the generation of harmful content. For example, \citet{meade-etal-2023-using} propose a method that retrieves safe response examples from similar dialogue contexts when generating responses in unsafe dialogue contexts, and uses these examples as in-context learning references.
In addition, numerous other studies have been conducted in this area, including surveys such as one by \citet{dong-etal-2024-attacks}. The aforementioned "guardrails" also perform tasks like fact-checking and hallucination detection. Furthermore, safety includes not engaging in unethical behavior. Many studies have also explored the ethical issues of dialogue systems \cite{henderson-ethical-challenges}.

Privacy protection is particularly important because users may easily disclose personal information during conversations. Therefore, it is essential to manage dialogue data appropriately. \citet{gumesel-privacy} survey privacy concerns from a socio-informatics perspective, while \citet{FAZZINGA2022200113} propose an architecture that enables privacy protection.

\paragraph{Unexplored Issues}

Discussions on AI safety are progressing, and various safety and ethical guidelines have been established \cite{jobin2019global, ISO/IEC42001:2023, IEEE7000:2021}. Research on methodologies to comply with these guidelines and with the EU AI Act is desirable.

\subsection{Dialogue Systems Engineering Professional Practice} \label{pracice}

Software engineering professional practice refers to the knowledge, skills, and attitudes that software engineers must possess to perform their work in a professional, responsible, and ethical manner. Dialogue system engineers are similarly expected to possess domain-specific knowledge, skills, and professional attitudes related to dialogue system technologies.

Although SWEBOK does not extensively address education in software engineering knowledge, education is particularly important for dialogue system engineers, as the field requires expertise in a wide range of technologies and knowledge. Several dialogue system development tools have been designed to help learners acquire such skills: for example, Advisor \cite{ortega-etal-2019-adviser} and DialBB \cite{nakano-komatani-2024-dialbb}.

In addition, various competitions and challenges are available to support the learning of dialogue system technologies. The Dialogue State Tracking Challenge (DSTC) \cite{williams2016dialog}, later rebranded as the Dialogue System Technology Challenge \cite{10174647}, began with a focus on dialogue state tracking and has since expanded to cover a wide range of dialogue system tasks. In contrast to DSTC's offline evaluation, the Alexa Prize \cite{https://doi.org/10.1609/aimag.v39i3.2810} is a competition in which systems are evaluated through interactions with real users. Other competitions with online evaluation formats include the Dialogue System Live Competition \cite{rh:iwsds24} and the Dialogue Robot Competition \cite{minato:ar23}. Participation in these competitions and challenges provides practical opportunities to learn and apply dialogue system technologies.

\paragraph{Unexplored Issues}

Although SWEBOK addresses the ethical attitudes required of engineers, we did not identify any studies that specifically address the ethical attitudes unique to dialogue system engineers. Nevertheless, ethical issues are inherent in the research and development of dialogue systems. In addition to ensuring that the dialogue system itself behaves ethically, the development process must also adhere to ethical standards. For example, ethical considerations must be taken into account during data collection and system evaluation. While such issues are shared with other fields in AI and HCI, dialogue systems involve direct interaction with humans through language, which requires particular attention.

\subsection{Dialogue System Economics} \label{economics}

Software engineering economics is a field that analyzes and evaluates the economic aspects of software development, including cost, value, risk, and profit. It aims to support decision-making in software engineering from an economic perspective, considering factors such as {\em cost-effectiveness}, {\em return on investment} (ROI), and {\em risk-benefit trade-offs}.

Aforementioned work by \citet{nakano:iwsds25} presents a methodology for listing evaluation items of dialogue systems from various perspectives, including economic ones.

\paragraph{Unexplored Issues}

There has been very little research analyzing the economic aspects of dialogue system development and operation. Before the advent of LLMs, the development of commercial dialogue systems was primarily undertaken by IT giants and specialized startups. Moving forward, more companies are likely to enter a phase where development proceeds only after careful assessment of business viability. For the broader adoption of dialogue systems, the development of methodologies for economic analysis is highly desirable.

\section{Future Directions}

In the previous section, we surveyed each knowledge area of dialogue systems engineering and discussed the remaining challenges. In this section, we consider the overall future directions for the field.

\paragraph{Methodologies for requirement specification and quality attributes}

There is currently a lack of research on methodologies for defining requirement specifications and quality attributes, particularly for the ISO/IEC 25010 characteristics of compatibility, reliability, maintainability, and portability. These qualities are essential for the reliable operation of dialogue systems. For example, in multimodal dialogue systems that transmit audio and images to servers, failing to include reliability as a quality requirement may lead to oversight during the design phase, especially when the system is deployed in environments with limited network bandwidth. Moreover, cost-benefit considerations should also be included in the formulation of specifications and quality attributes, highlighting the need for methodologies that incorporate economic perspectives.

\paragraph{Evaluation of architecture and design}

With the advancement of LLMs, the architecture and design of dialogue systems are undergoing significant changes. For instance, to mitigate {\em hallucinations} in LLM-based response generation \cite{nlg-hallucination-survey}, hybrid architectures that incorporate rule-based or example-based generation are becoming more common. However, as noted in the previous section, dialogue system architectures and designs are rarely evaluated from a software engineering perspective. Applying software engineering metrics such as cohesion and coupling to evaluate architectures would support the selection of appropriate architectures based on the application domain and use context.

\paragraph{Best practices for deployment, operation, maintenance, and continuous improvement}

Compared to design, construction, and testing, there is a noticeable lack of research on methodologies for deployment, operation, maintenance, and continuous improvement. For dialogue systems to be widely adopted in practical settings, it would be beneficial to share best practices in these areas. Additionally, sharing lessons learned from failed deployments or maintenance efforts could contribute valuable insights.

\paragraph{Tools supporting the entire life cycle}

Since dialogue systems require the integration of various technologies, tasks in each phase of the life cycle tend to be complex. To facilitate these tasks, an integrated ecosystem that supports design, construction, testing, deployment, operation, and continuous improvement would be highly beneficial. The availability of customizable and open tools for such an ecosystem would promote research across the full range of topics in the dialogue system life cycle.

\section{Concluding Remarks}

This paper referred to the field of software engineering related to dialogue systems  
as dialogue systems engineering and enumerated its knowledge areas based on the SWEBOK knowledge areas. We conducted a survey of existing research for each of these areas and, based on the results, discussed future directions.

Dialogue systems engineering is already being practiced in various ways in the development of research and commercial systems and frameworks. Sharing such practices as open and abstract knowledge can contribute to future research and development in dialogue systems.

One of the challenges, however, is that quantitative evaluation in dialogue systems engineering is difficult, making it hard to share knowledge in academia. For instance, it seems impractical to evaluate tools by comparing the development effort of systems built by engineers with similar skill levels, where the only difference is in specific tool features. Yet this does not mean that those features lack value. Therefore, it might be necessary to qualitatively evaluate such functionalities in the short term, while evaluating them through repeated use in the development of diverse systems in the long term.

Research related to dialogue systems engineering is being published not only in the communities of dialogue systems, natural language processing, and AI, but also in software engineering, healthcare, and other domains. It is essential for dialogue systems researchers to collaborate with these research communities to promote the development of practical systems. We hope that this paper serves as a first step in that direction.

\section*{Limitations}

As noted above, the survey presented here is not a systematic survey and therefore is not comprehensive. Consequently, there is a possibility that research already exists on some of the challenges we identified as underexplored. We will conduct a more comprehensive survey in the future.

Dialogue systems engineering is practiced in the development and operation of commercial systems, as well as in the tools used to build such systems. Many companies provide tools for building, deploying, and managing dialogue systems. Software engineering approaches in these commercial systems and tools are sometimes described in user manuals, technical blogs, and white papers.
In this paper, as a first step, we focused our investigation on academic papers, aiming to identify areas that have not yet been shared or systematized within the academic community and to uncover new research challenges.
However, bridging the gap between academia and industry is crucial for the future of dialogue systems research. As done in the software engineering community, it is necessary to establish a cycle in which industrial practices are systematized in academia and then fed back into industry. As the next step, we plan to include industrial practices in our investigation.

\bibliography{custom}

\begin{thebibliography}{138}
\providecommand{\natexlab}[1]{#1}

\bibitem[{Afzal et~al.(2019)Afzal, Dhamecha, Mukhi, Sindhgatta, Marvaniya,
  Ventura, and Yarbro}]{afzal-etal-2019-development}
Shazia Afzal, Tejas Dhamecha, Nirmal Mukhi, Renuka Sindhgatta, Smit Marvaniya,
  Matthew Ventura, and Jessica Yarbro. 2019.
\newblock \href {https://doi.org/10.18653/v1/N19-2015} {Development and
  deployment of a large-scale dialog-based intelligent tutoring system}.
\newblock In \emph{Proceedings of the 2019 Conference of the North {A}merican
  Chapter of the Association for Computational Linguistics: Human Language
  Technologies, Volume 2 (Industry Papers)}, pages 114--121, Minneapolis,
  Minnesota. Association for Computational Linguistics.

\bibitem[{Ai and Litman(2009)}]{ai-litman-2009-setting}
Hua Ai and Diane Litman. 2009.
\newblock \href {https://aclanthology.org/P09-1100/} {Setting up user action
  probabilities in user simulations for dialog system development}.
\newblock In \emph{Proceedings of the Joint Conference of the 47th Annual
  Meeting of the {ACL} and the 4th International Joint Conference on Natural
  Language Processing of the {AFNLP}}, pages 888--896, Suntec, Singapore.
  Association for Computational Linguistics.

\bibitem[{Ai et~al.(2007)Ai, Raux, Bohus, Eskenazi, and
  Litman}]{ai-etal-2007-comparing}
Hua Ai, Antoine Raux, Dan Bohus, Maxine Eskenazi, and Diane Litman. 2007.
\newblock \href {https://aclanthology.org/2007.sigdial-1.23/} {Comparing spoken
  dialog corpora collected with recruited subjects versus real users}.
\newblock In \emph{Proceedings of the 8th SIGdial Workshop on Discourse and
  Dialogue}, pages 124--131, Antwerp, Belgium. Association for Computational
  Linguistics.

\bibitem[{Aicher et~al.(2023)Aicher, Hillmann, Feustel, Michael, M\"{o}ller,
  and Minker}]{aicher:iwsds23}
Annalena Aicher, Stefan Hillmann, Isabel Feustel, Thilo Michael, Sebastian
  M\"{o}ller, and Wolfgang Minker. 2023.
\newblock \href {https://arxiv.org/abs/2411.11137v1} {Factors in crowdsourcing
  for evaluation of complex dialogue systems}.
\newblock In \emph{Proceedings of the 13th International Workshop on Spoken
  Dialogue Systems Technology (IWSDS 2023)}.

\bibitem[{Algherairy and Ahmed(2025)}]{ALGHERAIRY2025101697}
Atheer Algherairy and Moataz Ahmed. 2025.
\newblock \href {https://doi.org/10.1016/j.csl.2024.101697} {Prompting large
  language models for user simulation in task-oriented dialogue systems}.
\newblock \emph{Computer Speech \& Language}, 89:101697.

\bibitem[{Allen et~al.(2000)Allen, Byron, Dzikovska, Ferguson, Galescu, and
  Stent}]{allen:nle00}
James Allen, Donna Byron, Myroslava Dzikovska, George Ferguson, Lucian Galescu,
  and Amanda Stent. 2000.
\newblock \href {https://doi.org/10.1017/S135132490000245X} {An architecture
  for a generic dialogue shell}.
\newblock \emph{Natural Language Engineering}, 6(3–4):213--228.

\bibitem[{Allen et~al.(1995)Allen, Schubert, Ferguson, Heeman, Hwang, Kato,
  Light, Martin, Miller, Poesio, and Traum}]{ALLEN01011995}
James~F. Allen, Lenhart~K. Schubert, George Ferguson, Peter~A. Heeman,
  Chung~Hee Hwang, Tsuneaki Kato, Marc Light, Nathaniel~G. Martin, Bradford~W.
  Miller, Massimo Poesio, and David~R. Traum. 1995.
\newblock \href {https://doi.org/10.1080/09528139508953799} {The trains
  project: a case study in building a conversational planning agent}.
\newblock \emph{Journal of Experimental \& Theoretical Artificial
  Intelligence}, 7(1):7--48.

\bibitem[{Amershi et~al.(2019)Amershi, Begel, Bird, DeLine, Gall, Kamar,
  Nagappan, Nushi, and Zimmermann}]{8804457}
Saleema Amershi, Andrew Begel, Christian Bird, Robert DeLine, Harald Gall, Ece
  Kamar, Nachiappan Nagappan, Besmira Nushi, and Thomas Zimmermann. 2019.
\newblock \href {https://doi.org/10.1109/ICSE-SEIP.2019.00042} {Software
  engineering for machine learning: A case study}.
\newblock In \emph{2019 IEEE/ACM 41st International Conference on Software
  Engineering: Software Engineering in Practice (ICSE-SEIP)}, pages 291--300.

\bibitem[{Andrist et~al.(2017)Andrist, Bohus, Kamar, and Horvitz}]{Andrist17}
Sean Andrist, Dan Bohus, Ece Kamar, and Eric Horvitz. 2017.
\newblock \href
  {https://link.springer.com/chapter/10.1007/978-3-319-70022-9_29} {What went
  wrong and why? diagnosing situated interaction failures in the wild}.
\newblock In \emph{Social Robotics}, pages 293--303, Cham. Springer
  International Publishing.

\bibitem[{Araki(2012)}]{araki-2012-rapid}
Masahiro Araki. 2012.
\newblock \href {https://aclanthology.org/W12-1608/} {Rapid development process
  of spoken dialogue systems using collaboratively constructed semantic
  resources}.
\newblock In \emph{Proceedings of the 13th Annual Meeting of the Special
  Interest Group on Discourse and Dialogue}, pages 70--73, Seoul, South Korea.
  Association for Computational Linguistics.

\bibitem[{Atefi and
  Alipour(2019)}]{atefi2019automatedtestingframeworkconversational}
Soodeh Atefi and Mohammad~Amin Alipour. 2019.
\newblock \href {https://arxiv.org/abs/1902.06193} {An automated testing
  framework for conversational agents}.
\newblock \emph{Preprint}, arXiv:1902.06193.

\bibitem[{Atwell et~al.(2000)Atwell, Howarth, Souter, Baldo, Bisiani, Pezzotta,
  Bonaventura, Menzel, Herrorn, MortnoN, and et~al.}]{ATWEL2000}
Eric Atwell, Peter Howarth, Clive Souter, Patrizio Baldo, Roberto Bisiani,
  Dario Pezzotta, Patrizia Bonaventura, Wolfgang Menzel, DanielL Herrorn,
  Rachel MortnoN, and et~al. 2000.
\newblock \href {https://doi.org/10.1017/S1351324900002473} {User-guided system
  development in interactive spoken language education}.
\newblock \emph{Natural Language Engineering}, 6(3--4):229--241.

\bibitem[{Bocklisch et~al.(2017)Bocklisch, Faulkner, Pawlowski, and
  Nichol}]{bocklisch2017rasaopensourcelanguage}
Tom Bocklisch, Joey Faulkner, Nick Pawlowski, and Alan Nichol. 2017.
\newblock \href {https://arxiv.org/abs/1712.05181} {Rasa: Open source language
  understanding and dialogue management}.
\newblock \emph{Preprint}, arXiv:1712.05181.

\bibitem[{Bohus et~al.(2021)Bohus, Andrist, Feniello, Saw, Jalobeanu, Sweeney,
  Thompson, and Horvitz}]{bohus:arxiv21}
Dan Bohus, Sean Andrist, Ashley Feniello, Nick Saw, Mihai Jalobeanu, Patrick
  Sweeney, Anne~Loomis Thompson, and Eric Horvitz. 2021.
\newblock \href {https://arxiv.org/abs/2103.15975} {Platform for situated
  intelligence}.
\newblock \emph{CoRR}, abs/2103.15975.

\bibitem[{Bohus et~al.(2007)Bohus, Raux, Harris, Eskenazi, and
  Rudnicky}]{bohus-etal-2007-olympus}
Dan Bohus, Antoine Raux, Thomas Harris, Maxine Eskenazi, and Alexander
  Rudnicky. 2007.
\newblock \href {https://aclanthology.org/W07-0305/} {{O}lympus: an open-source
  framework for conversational spoken language interface research}.
\newblock In \emph{Proceedings of the Workshop on Bridging the Gap: Academic
  and Industrial Research in Dialog Technologies}, pages 32--39, Rochester, NY.
  Association for Computational Linguistics.

\bibitem[{Carrera-Rivera et~al.(2022)Carrera-Rivera, Ochoa, Larrinaga, and
  Lasa}]{CARRERARIVERA2022101895}
Angela Carrera-Rivera, William Ochoa, Felix Larrinaga, and Ganix Lasa. 2022.
\newblock \href {https://doi.org/10.1016/j.mex.2022.101895} {How-to conduct a
  systematic literature review: A quick guide for computer science research}.
\newblock \emph{MethodsX}, 9:101895.

\bibitem[{Chen et~al.(2020)Chen, Cheung, and
  Yiu}]{chen2020metamorphictestingnewapproach}
T.~Y. Chen, S.~C. Cheung, and S.~M. Yiu. 2020.
\newblock \href {https://arxiv.org/abs/2002.12543} {Metamorphic testing: A new
  approach for generating next test cases}.
\newblock \emph{Preprint}, arXiv:2002.12543.

\bibitem[{Chiba et~al.(2024)Chiba, Mitsuda, Lee, and
  Higashinaka}]{chiba:iwsds24}
Yuya Chiba, Koh Mitsuda, Akinobu Lee, and Ryuichiro Higashinaka. 2024.
\newblock The remdis toolkit: Building advanced real-time multimodal dialogue
  systems with incremental processing and large language models.
\newblock In \emph{Proceedings of the 14th International Workshop on Spoken
  Dialogue Systems Technology (IWSDS 2024)}.

\bibitem[{Chung et~al.(2023)Chung, Cahyawijaya, Wilie, Lovenia, and
  Fung}]{chung2023instructtodslargelanguagemodels}
Willy Chung, Samuel Cahyawijaya, Bryan Wilie, Holy Lovenia, and Pascale Fung.
  2023.
\newblock \href {https://arxiv.org/abs/2310.08885} {Instructtods: Large
  language models for end-to-end task-oriented dialogue systems}.
\newblock \emph{Preprint}, arXiv:2310.08885.

\bibitem[{Clark et~al.(2019)Clark, Pantidi, Cooney, Doyle, Garaialde, Edwards,
  Spillane, Gilmartin, Murad, Munteanu, Wade, and Cowan}]{clark:chi19}
Leigh Clark, Nadia Pantidi, Orla Cooney, Philip Doyle, Diego Garaialde, Justin
  Edwards, Brendan Spillane, Emer Gilmartin, Christine Murad, Cosmin Munteanu,
  Vincent Wade, and Benjamin~R. Cowan. 2019.
\newblock \href {https://doi.org/10.1145/3290605.3300705} {What makes a good
  conversation? {Challenges} in designing truly conversational agents}.
\newblock In \emph{Proceedings of the 2019 CHI Conference on Human Factors in
  Computing Systems}, CHI '19, page 1–12, New York, NY, USA. Association for
  Computing Machinery.

\bibitem[{Dahlb{\"a}ck and J{\"o}nsson(2003)}]{dahlback2003experiences}
Nils Dahlb{\"a}ck and Arne J{\"o}nsson. 2003.
\newblock \href
  {https://liu.diva-portal.org/smash/record.jsf?dswid=-9451&pid=diva2%3A355086}
  {Experiences with and lessons learned from working with a modular natural
  language dialogue architecture}.
\newblock In \emph{The 10th International Conference on Human-Computer
  Interaction, Crete}. Lawrence Erlbaum Associates.

\bibitem[{Daniel et~al.(2019)Daniel, Cabot, Deruelle, and Derras}]{Jarvis}
Gwendal Daniel, Jordi Cabot, Laurent Deruelle, and Mustapha Derras. 2019.
\newblock \href
  {https://link.springer.com/chapter/10.1007/978-3-030-21290-2_12}
  {Multi-platform chatbot modeling and deployment with the {Jarvis} framework}.
\newblock In \emph{Advanced Information Systems Engineering}, pages 177--193,
  Cham. Springer International Publishing.

\bibitem[{Daniel et~al.(2020)Daniel, Cabot, Deruelle, and Derras}]{xatkit}
Gwendal Daniel, Jordi Cabot, Laurent Deruelle, and Mustapha Derras. 2020.
\newblock \href {https://doi.org/10.1109/ACCESS.2020.2966919} {Xatkit: A
  multimodal low-code chatbot development framework}.
\newblock \emph{IEEE Access}, 8:15332--15346.

\bibitem[{Degerstedt and {J\"{o}nsson}(2006)}]{lintest}
Lars Degerstedt and Arne {J\"{o}nsson}. 2006.
\newblock \href {https://doi.org/10.21437/Interspeech.2006-154} {{LINTest}: a
  development tool for testing dialogue systems}.
\newblock In \emph{Proceedings of Interspeech 2006}, pages 225--235.

\bibitem[{Denecke(2002)}]{denecke-2002-rapid}
Matthias Denecke. 2002.
\newblock \href {https://aclanthology.org/C02-1147/} {Rapid prototyping for
  spoken dialogue systems}.
\newblock In \emph{{COLING} 2002: The 19th International Conference on
  Computational Linguistics}.

\bibitem[{Dong et~al.(2024)Dong, Zhou, Yang, Shao, and
  Qiao}]{dong-etal-2024-attacks}
Zhichen Dong, Zhanhui Zhou, Chao Yang, Jing Shao, and Yu~Qiao. 2024.
\newblock \href {https://doi.org/10.18653/v1/2024.naacl-long.375} {Attacks,
  defenses and evaluations for {LLM} conversation safety: A survey}.
\newblock In \emph{Proceedings of the 2024 Conference of the North American
  Chapter of the Association for Computational Linguistics: Human Language
  Technologies (Volume 1: Long Papers)}, pages 6734--6747, Mexico City, Mexico.
  Association for Computational Linguistics.

\bibitem[{Dybkj{\ae}r and Bernsen(2002)}]{dybkjaer2002dialogue}
Laila Dybkj{\ae}r and Niels~Ole Bernsen. 2002.
\newblock \href
  {http://spokendialogue.dk/Publications/2002i/LCmodel-23.10.2001-F-formatted2002.pdf}
  {The dialogue engineering life-cycle}.
\newblock \emph{Publications of the Department of General Linguistics,
  University of Tartu}, 3:103--125.

\bibitem[{Défossez et~al.(2024)Défossez, Mazaré, Orsini, Royer, Pérez,
  Jégou, Grave, and Zeghidour}]{defossez2024moshispeechtextfoundationmodel}
Alexandre Défossez, Laurent Mazaré, Manu Orsini, Amélie Royer, Patrick
  Pérez, Hervé Jégou, Edouard Grave, and Neil Zeghidour. 2024.
\newblock \href {https://arxiv.org/abs/2410.00037} {Moshi: a speech-text
  foundation model for real-time dialogue}.
\newblock \emph{Preprint}, arXiv:2410.00037.

\bibitem[{Esk{\'{e}}nazi et~al.(2008)Esk{\'{e}}nazi, Black, Raux, and
  Langner}]{lets-go-lab}
Maxine Esk{\'{e}}nazi, Alan~W. Black, Antoine Raux, and Brian Langner. 2008.
\newblock \href
  {https://www.isca-speech.org/archive/interspeech\_2008/eskenazi08\_interspeech.html}
  {Let's go lab: a platform for evaluation of spoken dialog systems with real
  world users}.
\newblock In \emph{9th Annual Conference of the International Speech
  Communication Association, {INTERSPEECH} 2008, Brisbane, Australia, September
  22-26, 2008}, page 219. {ISCA}.

\bibitem[{Farquhar et~al.(2024)Farquhar, Kossen, Kuhn, and
  Gal}]{farquhar2024detecting}
Sebastian Farquhar, Jannik Kossen, Lorenz Kuhn, and Yarin Gal. 2024.
\newblock \href {https://www.nature.com/articles/s41586-024-07421-0} {Detecting
  hallucinations in large language models using semantic entropy}.
\newblock \emph{Nature}, 630(8017):625--630.

\bibitem[{Fazzinga et~al.(2022)Fazzinga, Galassi, and
  Torroni}]{FAZZINGA2022200113}
Bettina Fazzinga, Andrea Galassi, and Paolo Torroni. 2022.
\newblock \href {https://doi.org/10.1016/j.iswa.2022.200113} {A
  privacy-preserving dialogue system based on argumentation}.
\newblock \emph{Intelligent Systems with Applications}, 16:200113.

\bibitem[{Ferguson and Allen(1998)}]{DBLP:conf/aaai/FergusonA98}
George Ferguson and James~F. Allen. 1998.
\newblock \href
  {https://aaai.org/papers/00567-aaai98-080-trips-an-integrated-intelligent-problem-solving-assistant/}
  {{TRIPS:} an integrated intelligent problem-solving assistant}.
\newblock In \emph{Proceedings of the Fifteenth National Conference on
  Artificial Intelligence and Tenth Innovative Applications of Artificial
  Intelligence Conference, {AAAI} 98, {IAAI} 98, July 26-30, 1998, Madison,
  Wisconsin, {USA}}, pages 567--572. {AAAI} Press / The {MIT} Press.

\bibitem[{Finch et~al.(2023)Finch, Paek, and Choi}]{finch-etal-2023-leveraging}
Sarah~E. Finch, Ellie~S. Paek, and Jinho~D. Choi. 2023.
\newblock \href {https://doi.org/10.18653/v1/2023.sigdial-1.20} {Leveraging
  large language models for automated dialogue analysis}.
\newblock In \emph{Proceedings of the 24th Annual Meeting of the Special
  Interest Group on Discourse and Dialogue}, pages 202--215, Prague, Czechia.
  Association for Computational Linguistics.

\bibitem[{Flycht-Eriksson and
  Jonsson(2000)}]{flycht-eriksson-jonsson-2000-dialogue}
Annika Flycht-Eriksson and Arne Jonsson. 2000.
\newblock \href {https://doi.org/10.3115/1117736.1117750} {Dialogue and domain
  knowledge management in dialogue systems}.
\newblock In \emph{1st {SIG}dial Workshop on Discourse and Dialogue}, pages
  121--130, Hong Kong, China. Association for Computational Linguistics.

\bibitem[{Fuchs et~al.(2012)Fuchs, Tsourakis, and
  Rayner}]{fuchs-etal-2012-scalable}
Matthew Fuchs, Nikos Tsourakis, and Manny Rayner. 2012.
\newblock \href {https://aclanthology.org/L12-1226/} {A scalable architecture
  for web deployment of spoken dialogue systems}.
\newblock In \emph{Proceedings of the Eighth International Conference on
  Language Resources and Evaluation ({LREC}`12)}, pages 1309--1314, Istanbul,
  Turkey. European Language Resources Association (ELRA).

\bibitem[{Gao et~al.(2024)Gao, Xiong, Gao, Jia, Pan, Bi, Dai, Sun, Wang, and
  Wang}]{gao2024retrievalaugmentedgenerationlargelanguage}
Yunfan Gao, Yun Xiong, Xinyu Gao, Kangxiang Jia, Jinliu Pan, Yuxi Bi, Yi~Dai,
  Jiawei Sun, Meng Wang, and Haofen Wang. 2024.
\newblock \href {https://arxiv.org/abs/2312.10997} {Retrieval-augmented
  generation for large language models: A survey}.
\newblock \emph{Preprint}, arXiv:2312.10997.

\bibitem[{Glass and Weinstein(2001)}]{DBLP:conf/interspeech/GlassW01}
James~R. Glass and Eugene Weinstein. 2001.
\newblock \href {https://doi.org/10.21437/EUROSPEECH.2001-345} {Speechbuilder:
  facilitating spoken dialogue system development}.
\newblock In \emph{{EUROSPEECH} 2001 Scandinavia, 7th European Conference on
  Speech Communication and Technology, 2nd {INTERSPEECH} Event, Aalborg,
  Denmark, September 3-7, 2001}, pages 1335--1338. {ISCA}.

\bibitem[{G\'{o}mez-Abajo et~al.(2024)G\'{o}mez-Abajo, P\'{e}rez-Soler,
  Ca\~{n}izares, Guerra, and de~Lara}]{mutation-test}
Pablo G\'{o}mez-Abajo, Sara P\'{e}rez-Soler, Pablo~C. Ca\~{n}izares, Esther
  Guerra, and Juan de~Lara. 2024.
\newblock \href {https://doi.org/10.1145/3661167.3661220} {Mutation testing for
  task-oriented chatbots}.
\newblock In \emph{Proceedings of the 28th International Conference on
  Evaluation and Assessment in Software Engineering}, EASE '24, page 232–241,
  New York, NY, USA. Association for Computing Machinery.

\bibitem[{Gumusel(2024)}]{gumesel-privacy}
Ece Gumusel. 2024.
\newblock \href {https://doi.org/10.1002/asi.24898} {A literature review of
  user privacy concerns inconversational chatbots: A social informatics
  approach: Anannual review of information science and technology (arist)
  paper}.
\newblock \emph{Journal of the Association for Information Science and
  Technology}, 76(1):121--154.

\bibitem[{Guo et~al.(2024)Guo, Aleti, Neelofar, and
  Tantithamthavorn}]{guo2024mortarmetamorphicmultiturntesting}
Guoxiang Guo, Aldeida Aleti, Neelofar Neelofar, and Chakkrit Tantithamthavorn.
  2024.
\newblock \href {https://arxiv.org/abs/2412.15557} {Mortar: Metamorphic
  multi-turn testing for llm-based dialogue systems}.
\newblock \emph{Preprint}, arXiv:2412.15557.

\bibitem[{Hartholt et~al.(2022)Hartholt, Fast, Li, Kim, Leeds, and
  Mozgai}]{hartholt:iva22}
Arno Hartholt, Ed~Fast, Zongjian Li, Kevin Kim, Andrew Leeds, and Sharon
  Mozgai. 2022.
\newblock \href {https://doi.org/10.1145/3514197.3549671} {Re-architecting the
  virtual human toolkit: towards an interoperable platform for embodied
  conversational agent research and development}.
\newblock In \emph{Proceedings of the 22nd ACM International Conference on
  Intelligent Virtual Agents}, IVA '22, New York, NY, USA. Association for
  Computing Machinery.

\bibitem[{Hartholt et~al.(2013)Hartholt, Traum, Marsella, Shapiro, Stratou,
  Leuski, Morency, and Gratch}]{hartholt:iva13}
Arno Hartholt, David Traum, Stacy~C. Marsella, Ari Shapiro, Giota Stratou,
  Anton Leuski, Louis-Philippe Morency, and Jonathan Gratch. 2013.
\newblock \href
  {https://link.springer.com/chapter/10.1007/978-3-642-40415-3_33} {All
  together now}.
\newblock In \emph{Intelligent Virtual Agents}, pages 368--381, Berlin,
  Heidelberg. Springer Berlin Heidelberg.

\bibitem[{Hartikainen et~al.(2004)Hartikainen, Turunen, Hakulinen, Salonen, and
  Funk}]{hartikainen04_interspeech}
Mikko Hartikainen, Markku Turunen, Jaakko Hakulinen, Esa-Pekka Salonen, and
  J.~Adam Funk. 2004.
\newblock \href {https://doi.org/10.21437/Interspeech.2004-117} {Flexible
  dialogue management using distributed and dynamic dialogue control}.
\newblock In \emph{Interspeech 2004}, pages 197--200.

\bibitem[{Henderson et~al.(2018)Henderson, Sinha, Angelard-Gontier, Ke, Fried,
  Lowe, and Pineau}]{henderson-ethical-challenges}
Peter Henderson, Koustuv Sinha, Nicolas Angelard-Gontier, Nan~Rosemary Ke,
  Genevieve Fried, Ryan Lowe, and Joelle Pineau. 2018.
\newblock \href {https://doi.org/10.1145/3278721.3278777} {Ethical challenges
  in data-driven dialogue systems}.
\newblock In \emph{Proceedings of the 2018 AAAI/ACM Conference on AI, Ethics,
  and Society}, AIES '18, page 123–129, New York, NY, USA. Association for
  Computing Machinery.

\bibitem[{Herzog and Reithinger(2006)}]{Herzog2006}
Gerd Herzog and Norbert Reithinger. 2006.
\newblock \href {https://doi.org/10.1007/3-540-36678-4_4} {\emph{The SmartKom
  Architecture: A Framework for Multimodal Dialogue Systems}}, pages 55--70.
\newblock Springer Berlin Heidelberg, Berlin, Heidelberg.

\bibitem[{Higashinaka et~al.(2021)Higashinaka, D'Haro, Abu~Shawar, Banchs,
  Funakoshi, Inaba, Tsunomori, Takahashi, and Sedoc}]{Higashinaka2021}
Ryuichiro Higashinaka, Luis~F. D'Haro, Bayan Abu~Shawar, Rafael~E. Banchs,
  Kotaro Funakoshi, Michimasa Inaba, Yuiko Tsunomori, Tetsuro Takahashi, and
  Jo{\~a}o Sedoc. 2021.
\newblock \href {https://doi.org/10.1007/978-981-15-9323-9_38} {\emph{Overview
  of the Dialogue Breakdown Detection Challenge 4}}, pages 403--417.
\newblock Springer Singapore, Singapore.

\bibitem[{Higashinaka et~al.(2024)Higashinaka, Inaba, Qi, Sasaki, Funakoshi,
  Moriya, Sato, Minato, Sakai, Funayama, Komuro, Nishikawa, Makino, Kikuchi,
  and Usami}]{rh:iwsds24}
Ryuichiro Higashinaka, Michimasa Inaba, Zhiyang Qi, Yuta Sasaki, Kotrao
  Funakoshi, Shoji Moriya, Shiki Sato, Takashi Minato, Kurima Sakai, Tomo
  Funayama, Masato Komuro, Hiroyuki Nishikawa, Ryosaku Makino, Hirofumi
  Kikuchi, and Mayumi Usami. 2024.
\newblock \href {https://hdl.handle.net/11094/95314} {Dialogue system live
  competition goes multimodal: Analyzing the effects of multimodal information
  in situated dialogue systems}.
\newblock In \emph{Proceedings of the 14th International Workshop on Spoken
  Dialogue Systems Technology (IWSDS 2024)}.

\bibitem[{Honovich et~al.(2022)Honovich, Aharoni, Herzig, Taitelbaum,
  Kukliansy, Cohen, Scialom, Szpektor, Hassidim, and
  Matias}]{honovich-etal-2022-true-evaluating}
Or~Honovich, Roee Aharoni, Jonathan Herzig, Hagai Taitelbaum, Doron Kukliansy,
  Vered Cohen, Thomas Scialom, Idan Szpektor, Avinatan Hassidim, and Yossi
  Matias. 2022.
\newblock \href {https://doi.org/10.18653/v1/2022.naacl-main.287} {{TRUE}:
  Re-evaluating factual consistency evaluation}.
\newblock In \emph{Proceedings of the 2022 Conference of the North American
  Chapter of the Association for Computational Linguistics: Human Language
  Technologies}, pages 3905--3920, Seattle, United States. Association for
  Computational Linguistics.

\bibitem[{Hude{\v{c}}ek and Dusek(2023)}]{hudecek-dusek-2023-large}
Vojt{\v{e}}ch Hude{\v{c}}ek and Ondrej Dusek. 2023.
\newblock \href {https://doi.org/10.18653/v1/2023.sigdial-1.21} {Are large
  language models all you need for task-oriented dialogue?}
\newblock In \emph{Proceedings of the 24th Annual Meeting of the Special
  Interest Group on Discourse and Dialogue}, pages 216--228, Prague, Czechia.
  Association for Computational Linguistics.

\bibitem[{Huynh et~al.(2022{\natexlab{a}})Huynh, Chiang, Bigham, and
  Eskenazi}]{huynh-etal-2022-dialcrowd}
Jessica Huynh, Ting-Rui Chiang, Jeffrey Bigham, and Maxine Eskenazi.
  2022{\natexlab{a}}.
\newblock \href {https://aclanthology.org/2022.lrec-1.134/} {{D}ial{C}rowd 2.0:
  A quality-focused dialog system crowdsourcing toolkit}.
\newblock In \emph{Proceedings of the Thirteenth Language Resources and
  Evaluation Conference}, pages 1256--1263, Marseille, France. European
  Language Resources Association.

\bibitem[{Huynh et~al.(2022{\natexlab{b}})Huynh, Mehri, Jiao, and
  Eskenazi}]{huynh-etal-2022-dialport}
Jessica Huynh, Shikib Mehri, Cathy Jiao, and Maxine Eskenazi.
  2022{\natexlab{b}}.
\newblock \href {https://doi.org/10.18653/v1/2022.sigdial-1.11} {The
  {D}ial{P}ort tools}.
\newblock In \emph{Proceedings of the 23rd Annual Meeting of the Special
  Interest Group on Discourse and Dialogue}, pages 101--106, Edinburgh, UK.
  Association for Computational Linguistics.

\bibitem[{IEEE(2021)}]{IEEE7000:2021}
IEEE. 2021.
\newblock \href {https://ieeexplore.ieee.org/document/9536679} {Ieee standard
  model process for addressing ethical concerns during system design}.
\newblock \emph{IEEE Std 7000-2021}, pages 1--82.

\bibitem[{Ishikawa and Yoshioka(2019)}]{8836142}
Fuyuki Ishikawa and Nobukazu Yoshioka. 2019.
\newblock \href {https://doi.org/10.1109/CESSER-IP.2019.00009} {How do
  engineers perceive difficulties in engineering of machine-learning systems? -
  questionnaire survey}.
\newblock In \emph{2019 IEEE/ACM Joint 7th International Workshop on Conducting
  Empirical Studies in Industry (CESI) and 6th International Workshop on
  Software Engineering Research and Industrial Practice (SER\&IP)}, pages 2--9.

\bibitem[{Isobe et~al.(2003)Isobe, Hayakawa, Murao, Mizutani, Takeda, and
  Itakura}]{isobe:interspeech03}
Toshihiro Isobe, Shoji Hayakawa, Hiroya Murao, Tatsuji Mizutani, Kazuya Takeda,
  and Fumitada Itakura. 2003.
\newblock \href {https://doi.org/10.21437/EUROSPEECH.2003-272} {A study on
  domain recognition of spoken dialogue systems}.
\newblock In \emph{8th European Conference on Speech Communication and
  Technology, {EUROSPEECH} 2003 - {INTERSPEECH} 2003, Geneva, Switzerland,
  September 1-4, 2003}, pages 1889--1892. {ISCA}.

\bibitem[{ISO/IEC(2011)}]{ISO/IEC25010:2011}
ISO/IEC. 2011.
\newblock \href {https://www.iso.org/standard/35733.html} {{ISO/IEC} 25010:2011
  systems and software engineering — systems and software quality
  requirements and evaluation (square) — system and software quality models}.
\newblock (withdrawn).

\bibitem[{ISO/IEC(2023{\natexlab{a}})}]{ISO/IEC25010:2023}
ISO/IEC. 2023{\natexlab{a}}.
\newblock \href {https://www.iso.org/standard/78176.html} {{ISO/IEC} 25010:2023
  systems and software engineering — systems and software quality
  requirements and evaluation (square) — product quality model}.

\bibitem[{ISO/IEC(2023{\natexlab{b}})}]{ISO/IEC42001:2023}
ISO/IEC. 2023{\natexlab{b}}.
\newblock \href {https://www.iso.org/standard/42001} {{ISO/IEC} 42001:2023
  information technology — artificial intelligence — management system}.

\bibitem[{Jacovi et~al.(2020)Jacovi, El, Lavi, Boaz, Amid, Ronen, and
  Anaby-Tavor}]{jacovi2020improving}
Alon Jacovi, Ori~Bar El, Ofer Lavi, David Boaz, David Amid, Inbal Ronen, and
  Ateret Anaby-Tavor. 2020.
\newblock \href {https://ceur-ws.org/Vol-2666/KDD_Converse20_paper_3.pdf}
  {Improving task-oriented dialogue systems in production with conversation
  logs.}
\newblock In \emph{Converse@ KDD}.

\bibitem[{Ji et~al.(2023)Ji, Lee, Frieske, Yu, Su, Xu, Ishii, Bang, Madotto,
  and Fung}]{nlg-hallucination-survey}
Ziwei Ji, Nayeon Lee, Rita Frieske, Tiezheng Yu, Dan Su, Yan Xu, Etsuko Ishii,
  Ye~Jin Bang, Andrea Madotto, and Pascale Fung. 2023.
\newblock \href {https://doi.org/10.1145/3571730} {Survey of hallucination in
  natural language generation}.
\newblock \emph{ACM Computing Surveys}, 55(12).

\bibitem[{Jia and Harman(2011)}]{mutation-testing-survey}
Yue Jia and Mark Harman. 2011.
\newblock \href {https://doi.org/10.1109/TSE.2010.62} {An analysis and survey
  of the development of mutation testing}.
\newblock \emph{IEEE Transactions on Software Engineering}, 37(5):649--678.

\bibitem[{Jobin et~al.(2019)Jobin, Ienca, and Vayena}]{jobin2019global}
Anna Jobin, Marcello Ienca, and Effy Vayena. 2019.
\newblock \href {https://www.nature.com/articles/s42256-019-0088-2} {The global
  landscape of {AI} ethics guidelines}.
\newblock \emph{Nature machine intelligence}, 1(9):389--399.

\bibitem[{Johnston et~al.(2023)Johnston, Flagg, Gottardi, Sahai, Lu, Sagi, Dai,
  Goyal, Hedayatnia, Hu, Jin, Lange, Liu, Liu, Pressel, Shi, Yang, Zhang,
  Zhang, Ball, Bland, Hu, Ipek, Jeun, Rocker, Vaz, Iyengar, Liu, Mandal,
  Hakkani-Tür, and Ghanadan}]{Johnston2023}
Michael Johnston, Cris Flagg, Anna Gottardi, Sattvik Sahai, Yao Lu, Samyuth
  Sagi, Luke Dai, Prasoon Goyal, Behnam Hedayatnia, Lucy Hu, Di~Jin, Patrick
  Lange, Shaohua Liu, Sijia Liu, Daniel Pressel, Hangjie Shi, Zhejia Yang, Chao
  Zhang, Desheng Zhang, Leslie Ball, Kate Bland, Shui Hu, Osman Ipek, James
  Jeun, Heather Rocker, Lavina Vaz, Akshaya Iyengar, Yang Liu, Arindam Mandal,
  Dilek Hakkani-Tür, and Reza Ghanadan. 2023.
\newblock \href
  {https://www.amazon.science/alexa-prize/proceedings/advancing-open-domain-dialog-the-fifth-alexa-prize-socialbot-grand-challenge}
  {Advancing open domain dialog: The fifth {Alexa} {Prize} socialbot grand
  challenge}.
\newblock In \emph{Alexa Prize SocialBot Grand Challenge 5 Proceedings}.

\bibitem[{Jurc{\i}cek et~al.(2011)Jurc{\i}cek, Keizer, Ga{\v{s}}ic, Mairesse,
  Thomson, Yu, and Young}]{jurcicek2011real}
Filip Jurc{\i}cek, Simon Keizer, Milica Ga{\v{s}}ic, Francois Mairesse, Blaise
  Thomson, Kai Yu, and Steve Young. 2011.
\newblock \href
  {https://www.isca-archive.org/interspeech_2011/jurcicek11_interspeech.html#}
  {Real user evaluation of spoken dialogue systems using amazon mechanical
  turk}.
\newblock In \emph{Proceedings of INTERSPEECH}.

\bibitem[{Kang et~al.(2018)Kang, Zhang, Kummerfeld, Tang, and
  Mars}]{kang-etal-2018-data}
Yiping Kang, Yunqi Zhang, Jonathan~K. Kummerfeld, Lingjia Tang, and Jason Mars.
  2018.
\newblock \href {https://doi.org/10.18653/v1/N18-3005} {Data collection for
  dialogue system: A startup perspective}.
\newblock In \emph{Proceedings of the 2018 Conference of the North {A}merican
  Chapter of the Association for Computational Linguistics: Human Language
  Technologies, Volume 3 (Industry Papers)}, pages 33--40, New Orleans -
  Louisiana. Association for Computational Linguistics.

\bibitem[{Kawamoto et~al.(2004)Kawamoto, Shimodaira, Nitta, Nishimoto,
  Nakamura, Itou, Morishima, Yotsukura, Kai, Lee, Yamashita, Kobayashi, Tokuda,
  Hirose, Minematsu, Yamada, Den, Utsuro, and Sagayama}]{Kawamoto2004}
Shin-ichi Kawamoto, Hiroshi Shimodaira, Tsuneo Nitta, Takuya Nishimoto, Satoshi
  Nakamura, Katsunobu Itou, Shigeo Morishima, Tatsuo Yotsukura, Atsuhiko Kai,
  Akinobu Lee, Yoichi Yamashita, Takao Kobayashi, Keiichi Tokuda, Keikichi
  Hirose, Nobuaki Minematsu, Atsushi Yamada, Yasuharu Den, Takehito Utsuro, and
  Shigeki Sagayama. 2004.
\newblock \href {https://doi.org/10.1007/978-3-662-08373-4_9} {\emph{Galatea:
  Open-Source Software for Developing Anthropomorphic Spoken Dialog Agents}},
  pages 187--211.
\newblock Springer Berlin Heidelberg, Berlin, Heidelberg.

\bibitem[{Khatri et~al.(2018)Khatri, Venkatesh, Hedayatnia, Ram, Gabriel, and
  Prasad}]{https://doi.org/10.1609/aimag.v39i3.2810}
Chandra Khatri, Anu Venkatesh, Behnam Hedayatnia, Ashwin Ram, Raefer Gabriel,
  and Rohit Prasad. 2018.
\newblock \href {https://doi.org/10.1609/aimag.v39i3.2810} {Alexa prize —
  state of the art in conversational ai}.
\newblock \emph{AI Magazine}, 39(3):40--55.

\bibitem[{Komatani et~al.(2006)Komatani, Kanda, Nakano, Nakadai, Tsujino,
  Ogata, and Okuno}]{komatani-etal-2006-multi}
Kazunori Komatani, Naoyuki Kanda, Mikio Nakano, Kazuhiro Nakadai, Hiroshi
  Tsujino, Tetsuya Ogata, and Hiroshi~G. Okuno. 2006.
\newblock \href {https://aclanthology.org/W06-1302/} {Multi-domain spoken
  dialogue system with extensibility and robustness against speech recognition
  errors}.
\newblock In \emph{Proceedings of the 7th {SIG}dial Workshop on Discourse and
  Dialogue}, pages 9--17, Sydney, Australia. Association for Computational
  Linguistics.

\bibitem[{Komeili et~al.(2022)Komeili, Shuster, and
  Weston}]{komeili-etal-2022-internet}
Mojtaba Komeili, Kurt Shuster, and Jason Weston. 2022.
\newblock \href {https://doi.org/10.18653/v1/2022.acl-long.579}
  {{I}nternet-augmented dialogue generation}.
\newblock In \emph{Proceedings of the 60th Annual Meeting of the Association
  for Computational Linguistics (Volume 1: Long Papers)}, pages 8460--8478,
  Dublin, Ireland. Association for Computational Linguistics.

\bibitem[{Lala et~al.(2025)Lala, Elmers, Inoue, Pang, Ochi, and
  Kawahara}]{lala-etal-2025-scriptboard}
Divesh Lala, Mikey Elmers, Koji Inoue, Zi~Haur Pang, Keiko Ochi, and Tatsuya
  Kawahara. 2025.
\newblock \href {https://aclanthology.org/2025.iwsds-1.17/} {{S}cript{B}oard:
  Designing modern spoken dialogue systems through visual programming}.
\newblock In \emph{Proceedings of the 15th International Workshop on Spoken
  Dialogue Systems Technology}, pages 176--182, Bilbao, Spain. Association for
  Computational Linguistics.

\bibitem[{Larsson and Traum(2000)}]{LARSSON_TRAUM_2000}
Staffan Larsson and David~R. Traum. 2000.
\newblock \href {https://doi.org/10.1017/S1351324900002539} {Information state
  and dialogue management in the {TRINDI} dialogue move engine toolkit}.
\newblock \emph{Natural Language Engineering}, 6(3--4):323--340.

\bibitem[{Lee et~al.(2013)Lee, Oura, and Tokuda}]{mmdagent}
Akinobu Lee, Keiichiro Oura, and Keiichi Tokuda. 2013.
\newblock \href {https://doi.org/10.1109/ICASSP.2013.6639300} {{MMDAgent}--a
  fully open-source toolkit for voice interaction systems}.
\newblock In \emph{Proc. ICASSP}, pages 8382--8385.

\bibitem[{Lee et~al.(2009)Lee, Jung, Kim, and Lee}]{LEE2009466}
Cheongjae Lee, Sangkeun Jung, Seokhwan Kim, and Gary~Geunbae Lee. 2009.
\newblock \href {https://doi.org/10.1016/j.specom.2009.01.008} {Example-based
  dialog modeling for practical multi-domain dialog system}.
\newblock \emph{Speech Communication}, 51(5):466--484.

\bibitem[{Lee et~al.(2017)Lee, Zhao, Du, Cai, Lu, Pincus, Traum, Ultes,
  Rojas-Barahona, Gasic, Young, and Eskenazi}]{lee-etal-2017-dialport}
Kyusong Lee, Tiancheng Zhao, Yulun Du, Edward Cai, Allen Lu, Eli Pincus, David
  Traum, Stefan Ultes, Lina~M. Rojas-Barahona, Milica Gasic, Steve Young, and
  Maxine Eskenazi. 2017.
\newblock \href {https://doi.org/10.18653/v1/W17-5521} {{D}ial{P}ort, gone
  live: An update after a year of development}.
\newblock In \emph{Proceedings of the 18th Annual {SIG}dial Meeting on
  Discourse and Dialogue}, pages 170--173, Saarbr{\"u}cken, Germany.
  Association for Computational Linguistics.

\bibitem[{Leite et~al.(2019)Leite, Rocha, Kon, Milojicic, and
  Meirelles}]{devops:csur}
Leonardo Leite, Carla Rocha, Fabio Kon, Dejan Milojicic, and Paulo Meirelles.
  2019.
\newblock \href {https://doi.org/10.1145/3359981} {A survey of devops concepts
  and challenges}.
\newblock \emph{ACM Computing Surveys}, 52(6).

\bibitem[{Leuski and Artstein(2017)}]{leuski-artstein-2017-lessons}
Anton Leuski and Ron Artstein. 2017.
\newblock \href {https://doi.org/10.18653/v1/W17-5541} {Lessons in dialogue
  system deployment}.
\newblock In \emph{Proceedings of the 18th Annual {SIG}dial Meeting on
  Discourse and Dialogue}, pages 352--355, Saarbr{\"u}cken, Germany.
  Association for Computational Linguistics.

\bibitem[{Li et~al.(2022)Li, Tao, Gao, and Guo}]{li-review}
Xiaomin Li, Chuanqi Tao, Jerry Gao, and Hongjing Guo. 2022.
\newblock \href {https://doi.org/10.1109/AITest55621.2022.00021} {A review of
  quality assurance research of dialogue systems}.
\newblock In \emph{2022 IEEE International Conference On Artificial
  Intelligence Testing (AITest)}, pages 87--94.

\bibitem[{Li et~al.(2021)Li, Arnold, Yan, Shi, and Yu}]{li-etal-2021-legoeval}
Yu~Li, Josh Arnold, Feifan Yan, Weiyan Shi, and Zhou Yu. 2021.
\newblock \href {https://doi.org/10.18653/v1/2021.acl-demo.38} {{LEGOE}val: An
  open-source toolkit for dialogue system evaluation via crowdsourcing}.
\newblock In \emph{Proceedings of the 59th Annual Meeting of the Association
  for Computational Linguistics and the 11th International Joint Conference on
  Natural Language Processing: System Demonstrations}, pages 317--324, Online.
  Association for Computational Linguistics.

\bibitem[{Lison and Kennington(2016)}]{lison-kennington-2016-opendial}
Pierre Lison and Casey Kennington. 2016.
\newblock \href {https://doi.org/10.18653/v1/P16-4012} {{O}pen{D}ial: A toolkit
  for developing spoken dialogue systems with probabilistic rules}.
\newblock In \emph{Proceedings of {ACL}-2016 System Demonstrations}, pages
  67--72, Berlin, Germany. Association for Computational Linguistics.

\bibitem[{Liu et~al.(2024)Liu, Deng, Li, Wang, Wang, Wang, Zhang, Liu, Wang,
  Zheng, and Liu}]{liu2024promptinjectionattackllmintegrated}
Yi~Liu, Gelei Deng, Yuekang Li, Kailong Wang, Zihao Wang, Xiaofeng Wang,
  Tianwei Zhang, Yepang Liu, Haoyu Wang, Yan Zheng, and Yang Liu. 2024.
\newblock \href {https://arxiv.org/abs/2306.05499} {Prompt injection attack
  against {LLM}-integrated applications}.
\newblock \emph{Preprint}, arXiv:2306.05499.

\bibitem[{Lowe et~al.(2015)Lowe, Pow, Serban, and
  Pineau}]{lowe-etal-2015-ubuntu}
Ryan Lowe, Nissan Pow, Iulian Serban, and Joelle Pineau. 2015.
\newblock \href {https://doi.org/10.18653/v1/W15-4640} {The {U}buntu dialogue
  corpus: A large dataset for research in unstructured multi-turn dialogue
  systems}.
\newblock In \emph{Proceedings of the 16th Annual Meeting of the Special
  Interest Group on Discourse and Dialogue}, pages 285--294, Prague, Czech
  Republic. Association for Computational Linguistics.

\bibitem[{Luo et~al.(2024)Luo, Tang, Wang, and Zhang}]{luo-etal-2024-duetsim}
Xiang Luo, Zhiwen Tang, Jin Wang, and Xuejie Zhang. 2024.
\newblock \href {https://aclanthology.org/2024.lrec-main.481/} {{D}uet{S}im:
  Building user simulator with dual large language models for task-oriented
  dialogues}.
\newblock In \emph{Proceedings of the 2024 Joint International Conference on
  Computational Linguistics, Language Resources and Evaluation (LREC-COLING
  2024)}, pages 5414--5424, Torino, Italia. ELRA and ICCL.

\bibitem[{Madotto et~al.(2021)Madotto, Lin, Winata, and
  Fung}]{madotto2021fewshotbotpromptbasedlearning}
Andrea Madotto, Zhaojiang Lin, Genta~Indra Winata, and Pascale Fung. 2021.
\newblock \href {https://arxiv.org/abs/2110.08118} {Few-shot bot: Prompt-based
  learning for dialogue systems}.
\newblock \emph{Preprint}, arXiv:2110.08118.

\bibitem[{Manuvinakurike et~al.(2015)Manuvinakurike, Paetzel, and
  Devault}]{manuvinakurike-etal-2015-reducing}
Ramesh Manuvinakurike, Maike Paetzel, and David Devault. 2015.
\newblock \href
  {http://semdial.org/anthology/Z15-Manuvinakurike_semdial_0016.pdf} {Reducing
  the cost of dialogue system training and evaluation with online,
  crowd-sourced dialogue data collection}.
\newblock In \emph{Proceedings of the 19th Workshop on the Semantics and
  Pragmatics of Dialogue - Full Papers}, Gothenburg, Sweden. SEMDIAL.

\bibitem[{Mart\'{\i}nez-Fern\'{a}ndez et~al.(2022)Mart\'{\i}nez-Fern\'{a}ndez,
  Bogner, Franch, Oriol, Siebert, Trendowicz, Vollmer, and
  Wagner}]{10.1145/3487043}
Silverio Mart\'{\i}nez-Fern\'{a}ndez, Justus Bogner, Xavier Franch, Marc Oriol,
  Julien Siebert, Adam Trendowicz, Anna~Maria Vollmer, and Stefan Wagner. 2022.
\newblock \href {https://doi.org/10.1145/3487043} {Software engineering for
  ai-based systems: A survey}.
\newblock \emph{ACM Trans. Softw. Eng. Methodol.}, 31(2).

\bibitem[{McTear(2004)}]{McTear2004}
Michael~F. McTear. 2004.
\newblock \href {https://doi.org/10.1007/978-0-85729-414-2_7} {Dialogue
  engineering: The dialogue systems development lifecycle}.
\newblock In \emph{Spoken Dialogue Technology: Toward the Conversational User
  Interface}, pages 129--161. Springer London, London.

\bibitem[{Meade et~al.(2023)Meade, Gella, Hazarika, Gupta, Jin, Reddy, Liu, and
  Hakkani-Tur}]{meade-etal-2023-using}
Nicholas Meade, Spandana Gella, Devamanyu Hazarika, Prakhar Gupta, Di~Jin, Siva
  Reddy, Yang Liu, and Dilek Hakkani-Tur. 2023.
\newblock \href {https://doi.org/10.18653/v1/2023.findings-emnlp.796} {Using
  in-context learning to improve dialogue safety}.
\newblock In \emph{Findings of the Association for Computational Linguistics:
  EMNLP 2023}, pages 11882--11910, Singapore. Association for Computational
  Linguistics.

\bibitem[{Mehri and Eskenazi(2020)}]{mehri-eskenazi-2020-unsupervised}
Shikib Mehri and Maxine Eskenazi. 2020.
\newblock \href {https://doi.org/10.18653/v1/2020.sigdial-1.28} {Unsupervised
  evaluation of interactive dialog with {D}ialo{GPT}}.
\newblock In \emph{Proceedings of the 21th Annual Meeting of the Special
  Interest Group on Discourse and Dialogue}, pages 225--235, 1st virtual
  meeting. Association for Computational Linguistics.

\bibitem[{Mendes(2005)}]{web-engineering-review}
E.~Mendes. 2005.
\newblock \href {https://doi.org/10.1109/ISESE.2005.1541857} {A systematic
  review of web engineering research}.
\newblock In \emph{2005 International Symposium on Empirical Software
  Engineering, 2005.}

\bibitem[{Meng et~al.(2004)Meng, Ching, Chan, Wong, and Chan}]{meng:isis}
Helen Meng, P.~C. Ching, Shuk~Fong Chan, Yee~Fong Wong, and Cheong~Chat Chan.
  2004.
\newblock \href {https://doi.org/10.1145/1017494.1017497} {{ISIS}: an adaptive,
  trilingual conversational system with interleaving interaction and delegation
  dialogs}.
\newblock \emph{ACM Trans. Comput.-Hum. Interact.}, 11(3):268–299.

\bibitem[{Michael(2020)}]{michael-2020-retico}
Thilo Michael. 2020.
\newblock \href {https://doi.org/10.18653/v1/2020.sigdial-1.6} {Retico: An
  incremental framework for spoken dialogue systems}.
\newblock In \emph{Proceedings of the 21th Annual Meeting of the Special
  Interest Group on Discourse and Dialogue}, pages 49--52, 1st virtual meeting.
  Association for Computational Linguistics.

\bibitem[{Miller et~al.(2018)Miller, Feng, Fisch, Lu, Batra, Bordes, Parikh,
  and Weston}]{miller2018parlaidialogresearchsoftware}
Alexander~H. Miller, Will Feng, Adam Fisch, Jiasen Lu, Dhruv Batra, Antoine
  Bordes, Devi Parikh, and Jason Weston. 2018.
\newblock \href {https://arxiv.org/abs/1705.06476} {Parlai: A dialog research
  software platform}.
\newblock \emph{Preprint}, arXiv:1705.06476.

\bibitem[{Minato et~al.(2023)Minato, Higashinaka, Sakai, Funayama, Nishizaki,
  and Nagai}]{minato:ar23}
Takashi Minato, Ryuichiro Higashinaka, Kurima Sakai, Tomo Funayama, Hiromitsu
  Nishizaki, and Takayuki Nagai. 2023.
\newblock \href
  {https://www.tandfonline.com/doi/full/10.1080/01691864.2023.2249530} {Design
  of a competition specifically for spoken dialogue with a humanoid robot}.
\newblock \emph{Advanced Robotics}, 37(21):1349--1363.

\bibitem[{Motger et~al.(2022)Motger, Franch, and Marco}]{software-based-ds}
Quim Motger, Xavier Franch, and Jordi Marco. 2022.
\newblock \href {https://doi.org/10.1145/3527450} {Software-based dialogue
  systems: Survey, taxonomy, and challenges}.
\newblock \emph{ACM Computing Surveys}, 55(5).

\bibitem[{Nakano et~al.(2011)Nakano, Hasegawa, Funakoshi, Takeuchi, Torii,
  Nakadai, Kanda, Komatani, Okuno, and Tsujino}]{NAKANO2011248}
Mikio Nakano, Yuji Hasegawa, Kotaro Funakoshi, Johane Takeuchi, Toyotaka Torii,
  Kazuhiro Nakadai, Naoyuki Kanda, Kazunori Komatani, Hiroshi~G Okuno, and
  Hiroshi Tsujino. 2011.
\newblock \href {https://doi.org/10.1016/j.knosys.2010.08.004} {A multi-expert
  model for dialogue and behavior control of conversational robots and agents}.
\newblock \emph{Knowledge-Based Systems}, 24(2):248--256.

\bibitem[{Nakano and Komatani(2024)}]{nakano-komatani-2024-dialbb}
Mikio Nakano and Kazunori Komatani. 2024.
\newblock \href {https://doi.org/10.18653/v1/2024.sigdial-1.56} {{D}ial{BB}: A
  dialogue system development framework as an educational material}.
\newblock In \emph{Proceedings of the 25th Annual Meeting of the Special
  Interest Group on Discourse and Dialogue}, pages 664--668, Kyoto, Japan.
  Association for Computational Linguistics.

\bibitem[{Nakano et~al.(2025{\natexlab{a}})Nakano, Komatani, and
  Takeuchi}]{nakano-sigdial25}
Mikio Nakano, Kazunori Komatani, and Hironori Takeuchi. 2025{\natexlab{a}}.
\newblock Generating diverse personas for user simulators to test interview
  dialogue systems.
\newblock In \emph{Proceedings of the 26th Annual Meeting of the Special
  Interest Group on Discourse and Dialogue}, Avigon, France. Association for
  Computational Linguistics.
\newblock (to appear).

\bibitem[{Nakano et~al.(2024)Nakano, Mukai, Matsuyama, and
  Komatani}]{nakano:iwsds24}
Mikio Nakano, Hisahiro Mukai, Yoichi Matsuyama, and Kazunori Komatani. 2024.
\newblock \href {https://hdl.handle.net/11094/95314} {Evaluating dialogue
  systems from the system owners' perspectives}.
\newblock In \emph{Proceedings of the 14th International Workshop on Spoken
  Dialogue Systems Technology (IWSDS 2024)}.

\bibitem[{Nakano et~al.(2025{\natexlab{b}})Nakano, Takeuchi, and
  Komatani}]{nakano:iwsds25}
Mikio Nakano, Hironori Takeuchi, and Kazunori Komatani. 2025{\natexlab{b}}.
\newblock \href {https://aclanthology.org/2025.iwsds-1.24/} {A methodology for
  identifying evaluation items for practical dialogue systems based on
  business-dialogue system alignment models}.
\newblock In \emph{Proceedings of the 15th International Workshop on Spoken
  Dialogue Systems Technology (IWSDS 2025)}.

\bibitem[{Niu et~al.(2024)Niu, Wu, Zhu, Xu, Shum, Zhong, Song, and
  Zhang}]{niu-etal-2024-ragtruth}
Cheng Niu, Yuanhao Wu, Juno Zhu, Siliang Xu, KaShun Shum, Randy Zhong, Juntong
  Song, and Tong Zhang. 2024.
\newblock \href {https://doi.org/10.18653/v1/2024.acl-long.585} {{RAGT}ruth: A
  hallucination corpus for developing trustworthy retrieval-augmented language
  models}.
\newblock In \emph{Proceedings of the 62nd Annual Meeting of the Association
  for Computational Linguistics (Volume 1: Long Papers)}, pages 10862--10878,
  Bangkok, Thailand. Association for Computational Linguistics.

\bibitem[{Olabiyi et~al.(2020)Olabiyi, Bhattarai, Bruss, and
  Kulis}]{olabiyi2020dlgnettaskendtoendneuralnetwork}
Oluwatobi~O. Olabiyi, Prarthana Bhattarai, C.~Bayan Bruss, and Zachary Kulis.
  2020.
\newblock \href {https://arxiv.org/abs/2010.01693} {Dlgnet-task: An end-to-end
  neural network framework for modeling multi-turn multi-domain task-oriented
  dialogue}.
\newblock \emph{Preprint}, arXiv:2010.01693.

\bibitem[{O'Neill and McTear(2000)}]{ONEILL_McTEAR_2000}
Ian~M. O'Neill and Michael~F. McTear. 2000.
\newblock \href {https://doi.org/10.1017/S1351324900002527} {Object-oriented
  modelling of spoken language dialogue systems}.
\newblock \emph{Natural Language Engineering}, 6(3--4):341--362.

\bibitem[{Ortega et~al.(2019)Ortega, V{\"a}th, Weber, Vanderlyn, Schmidt,
  V{\"o}lkel, Karacevic, and Vu}]{ortega-etal-2019-adviser}
Daniel Ortega, Dirk V{\"a}th, Gianna Weber, Lindsey Vanderlyn, Maximilian
  Schmidt, Moritz V{\"o}lkel, Zorica Karacevic, and Ngoc~Thang Vu. 2019.
\newblock \href {https://doi.org/10.18653/v1/P19-3016} {{ADVISER}: A dialog
  system framework for education {\&} research}.
\newblock In \emph{Proceedings of the 57th Annual Meeting of the Association
  for Computational Linguistics: System Demonstrations}, pages 93--98,
  Florence, Italy. Association for Computational Linguistics.

\bibitem[{Pang et~al.(2024)Pang, Roller, Cho, He, and
  Weston}]{pang-etal-2024-leveraging}
Richard~Yuanzhe Pang, Stephen Roller, Kyunghyun Cho, He~He, and Jason Weston.
  2024.
\newblock \href {https://aclanthology.org/2024.eacl-short.8/} {Leveraging
  implicit feedback from deployment data in dialogue}.
\newblock In \emph{Proceedings of the 18th Conference of the European Chapter
  of the Association for Computational Linguistics (Volume 2: Short Papers)},
  pages 60--75, St. Julian{'}s, Malta. Association for Computational
  Linguistics.

\bibitem[{Pietquin and Hastie(2013)}]{pietquin2013survey}
Olivier Pietquin and Helen Hastie. 2013.
\newblock \href
  {https://www.cambridge.org/core/journals/knowledge-engineering-review/article/abs/survey-on-metrics-for-the-evaluation-of-user-simulations/602976EC6417B5BAA1719D0876FB5611}
  {A survey on metrics for the evaluation of user simulations}.
\newblock \emph{The knowledge engineering review}, 28(1):59--73.

\bibitem[{Ramanarayanan et~al.(2017)Ramanarayanan, Suendermann-Oeft, Molloy,
  Tsuprun, Lange, and Evanini}]{ramanarayanan2017crowdsourcing}
Vikram Ramanarayanan, David Suendermann-Oeft, Hillary Molloy, Eugene Tsuprun,
  Patrick~L Lange, and Keelan Evanini. 2017.
\newblock Crowdsourcing multimodal dialog interactions: Lessons learned from
  the halef case.
\newblock In \emph{The AAAI-17 Workshop on Crowdsourcing, Deep Learning, and
  Artificial Intelligence Agents}.

\bibitem[{Raux and Eskenazi(2007)}]{raux:asru2007}
Antoine Raux and Maxine Eskenazi. 2007.
\newblock \href {https://doi.org/10.1109/ASRU.2007.4430165} {A multi-layer
  architecture for semi-synchronous event-driven dialogue management}.
\newblock In \emph{2007 IEEE Workshop on Automatic Speech Recognition \&
  Understanding (ASRU)}, pages 514--519.

\bibitem[{Rebedea et~al.(2023)Rebedea, Dinu, Sreedhar, Parisien, and
  Cohen}]{rebedea-etal-2023-nemo}
Traian Rebedea, Razvan Dinu, Makesh~Narsimhan Sreedhar, Christopher Parisien,
  and Jonathan Cohen. 2023.
\newblock \href {https://doi.org/10.18653/v1/2023.emnlp-demo.40} {{N}e{M}o
  guardrails: A toolkit for controllable and safe {LLM} applications with
  programmable rails}.
\newblock In \emph{Proceedings of the 2023 Conference on Empirical Methods in
  Natural Language Processing: System Demonstrations}, pages 431--445,
  Singapore. Association for Computational Linguistics.

\bibitem[{Robino(2025)}]{robino2025conversationroutinespromptengineering}
Giorgio Robino. 2025.
\newblock \href {https://arxiv.org/abs/2501.11613} {Conversation routines: A
  prompt engineering framework for task-oriented dialog systems}.
\newblock \emph{Preprint}, arXiv:2501.11613.

\bibitem[{Roca et~al.(2020)Roca, Sancho, García, and Álvaro
  Alesanco}]{ROCA2020103305}
Surya Roca, Jorge Sancho, José García, and Álvaro Alesanco. 2020.
\newblock \href {https://doi.org/10.1016/j.jbi.2019.103305} {Microservice
  chatbot architecture for chronic patient support}.
\newblock \emph{Journal of Biomedical Informatics}, 102:103305.

\bibitem[{Schatzmann et~al.(2006)Schatzmann, Weilhammer, Stuttle, and
  Young}]{Schatzmann:survey}
Jost Schatzmann, Karl Weilhammer, Matt Stuttle, and Steve Young. 2006.
\newblock \href {https://doi.org/10.1017/S0269888906000944} {A survey of
  statistical user simulation techniques for reinforcement-learning of dialogue
  management strategies}.
\newblock \emph{Knowl. Eng. Rev.}, 21(2):97–126.

\bibitem[{Schl{\"{o}}gl et~al.(2014)Schl{\"{o}}gl, Doherty, and
  Luz}]{10.1093/iwc/iwu016}
Stephan Schl{\"{o}}gl, Gavin Doherty, and Saturnino Luz. 2014.
\newblock \href {https://doi.org/10.1093/iwc/iwu016} {Wizard of oz
  experimentation for language technology applications: Challenges and tools}.
\newblock \emph{Interacting with Computers}, 27(6):592--615.

\bibitem[{Sculley et~al.(2015)Sculley, Holt, Golovin, Davydov, Phillips, Ebner,
  Chaudhary, Young, Crespo, and Dennison}]{10.5555/2969442.2969519}
D.~Sculley, Gary Holt, Daniel Golovin, Eugene Davydov, Todd Phillips, Dietmar
  Ebner, Vinay Chaudhary, Michael Young, Jean-Francois Crespo, and Dan
  Dennison. 2015.
\newblock \href
  {https://papers.nips.cc/paper_files/paper/2015/hash/86df7dcfd896fcaf2674f757a2463eba-Abstract.html}
  {Hidden technical debt in machine learning systems}.
\newblock In \emph{Proceedings of the 29th International Conference on Neural
  Information Processing Systems - Volume 2}, NIPS'15, page 2503–2511,
  Cambridge, MA, USA. MIT Press.

\bibitem[{Seneff et~al.(1998)Seneff, Hurley, Lau, Pao, Schmid, and
  Zue}]{DBLP:conf/interspeech/SeneffHLPSZ98}
Stephanie Seneff, Edward Hurley, Raymond Lau, Christine Pao, Philipp Schmid,
  and Victor Zue. 1998.
\newblock \href {https://doi.org/10.21437/ICSLP.1998-478} {{GALAXY-II:} a
  reference architecture for conversational system development}.
\newblock In \emph{The 5th International Conference on Spoken Language
  Processing, Incorporating The 7th Australian International Speech Science and
  Technology Conference, Sydney Convention Centre, Sydney, Australia, 30th
  November - 4th December 1998}. {ISCA}.

\bibitem[{Skantze and Johansson(2015)}]{skantze-johansson-2015-modelling}
Gabriel Skantze and Martin Johansson. 2015.
\newblock \href {https://doi.org/10.18653/v1/W15-4624} {Modelling situated
  human-robot interaction using {I}ris{TK}}.
\newblock In \emph{Proceedings of the 16th Annual Meeting of the Special
  Interest Group on Discourse and Dialogue}, pages 165--167, Prague, Czech
  Republic. Association for Computational Linguistics.

\bibitem[{Song et~al.(2024)Song, Wang, Zhu, Wu, Cheng, Zhong, and
  Niu}]{song-etal-2024-rag}
Juntong Song, Xingguang Wang, Juno Zhu, Yuanhao Wu, Xuxin Cheng, Randy Zhong,
  and Cheng Niu. 2024.
\newblock \href {https://doi.org/10.18653/v1/2024.emnlp-industry.113}
  {{RAG}-{HAT}: A hallucination-aware tuning pipeline for {LLM} in
  retrieval-augmented generation}.
\newblock In \emph{Proceedings of the 2024 Conference on Empirical Methods in
  Natural Language Processing: Industry Track}, pages 1548--1558, Miami,
  Florida, US. Association for Computational Linguistics.

\bibitem[{Sonntag et~al.(2010)Sonntag, Reithinger, Herzog, and
  Becker}]{Sonntag:2010}
Daniel Sonntag, Norbert Reithinger, Gerd Herzog, and Tilman Becker. 2010.
\newblock \href
  {https://link.springer.com/chapter/10.1007/978-3-642-16202-2_12} {A discourse
  and dialogue infrastructure for industrial dissemination}.
\newblock In \emph{Spoken Dialogue Systems for Ambient Environments}, pages
  132--143, Berlin, Heidelberg. Springer Berlin Heidelberg.

\bibitem[{Stevens et~al.(1974)Stevens, Myers, and
  Constantine}]{structured-design}
W.~P. Stevens, G.~J. Myers, and L.~L. Constantine. 1974.
\newblock \href {https://doi.org/10.1147/sj.132.0115} {Structured design}.
\newblock \emph{IBM Systems Journal}, 13(2):115--139.

\bibitem[{Sutton et~al.(1996)Sutton, Novick, Cole, Vermeulen, de~Villiers,
  Schalkwyk, and Fanty}]{cslu-tookit}
S.~Sutton, D.G. Novick, R.~Cole, P.~Vermeulen, J.~de~Villiers, J.~Schalkwyk,
  and M.~Fanty. 1996.
\newblock \href {https://doi.org/10.1109/ICSLP.1996.607460} {Building 10,000
  spoken dialogue systems}.
\newblock In \emph{Proceeding of Fourth International Conference on Spoken
  Language Processing. ICSLP '96}, volume~2, pages 709--712 vol.2.

\bibitem[{Sutton et~al.(1998)Sutton, Cole, De~Villiers, Schalkwyk, Vermeulen,
  Macon, Yan, Kaiser, Rundle, Shobaki et~al.}]{sutton1998universal}
Stephen Sutton, Ronald~A Cole, Jacques De~Villiers, Johan Schalkwyk, Pieter~JE
  Vermeulen, Michael~W Macon, Yonghong Yan, Edward~C Kaiser, Brian Rundle,
  Khaldoun Shobaki, et~al. 1998.
\newblock \href {https://www.isca-archive.org/icslp_1998/sutton98_icslp.html}
  {Universal speech tools: the {CSLU} toolkit.}
\newblock In \emph{ICSLP}, volume~98, pages 3221--3224. Citeseer.

\bibitem[{Takanobu et~al.(2020)Takanobu, Zhu, Li, Peng, Gao, and
  Huang}]{takanobu-etal-2020-goal}
Ryuichi Takanobu, Qi~Zhu, Jinchao Li, Baolin Peng, Jianfeng Gao, and Minlie
  Huang. 2020.
\newblock \href {https://doi.org/10.18653/v1/2020.sigdial-1.37} {Is your
  goal-oriented dialog model performing really well? empirical analysis of
  system-wise evaluation}.
\newblock In \emph{Proceedings of the 21th Annual Meeting of the Special
  Interest Group on Discourse and Dialogue}, pages 297--310, 1st virtual
  meeting. Association for Computational Linguistics.

\bibitem[{Traum et~al.(2015)Traum, Jones, Hays, Maio, Alexander, Artstein,
  Debevec, Gainer, Georgila, Haase, Jungblut, Leuski, Smith, and
  Swartout}]{traum:ndt}
David Traum, Andrew Jones, Kia Hays, Heather Maio, Oleg Alexander, Ron
  Artstein, Paul Debevec, Alesia Gainer, Kallirroi Georgila, Kathleen Haase,
  Karen Jungblut, Anton Leuski, Stephen Smith, and William Swartout. 2015.
\newblock \href
  {https://link.springer.com/chapter/10.1007/978-3-319-27036-4_26} {New
  dimensions in testimony: Digitally preserving a holocaust survivor's
  interactive storytelling}.
\newblock In \emph{Interactive Storytelling}, pages 269--281, Cham. Springer
  International Publishing.

\bibitem[{Truong et~al.(2017)Truong, Parthasarathi, and
  Pineau}]{truong-etal-2017-maca}
Hoai~Phuoc Truong, Prasanna Parthasarathi, and Joelle Pineau. 2017.
\newblock \href {https://doi.org/10.18653/v1/W17-5513} {{MACA}: A modular
  architecture for conversational agents}.
\newblock In \emph{Proceedings of the 18th Annual {SIG}dial Meeting on
  Discourse and Dialogue}, pages 93--102, Saarbr{\"u}cken, Germany. Association
  for Computational Linguistics.

\bibitem[{Ultes and Maier(2021)}]{ultes:dd21}
Stefan Ultes and Wolfgang Maier. 2021.
\newblock \href {https://journals.uic.edu/ojs/index.php/dad/article/view/11518}
  {User satisfaction reward estimation across domains: Domain-independent
  dialogue policy learning}.
\newblock \emph{Dialogue and Discourse}, 12(2):81--114.

\bibitem[{Ultes et~al.(2017)Ultes, Rojas-Barahona, Su, Vandyke, Kim, Casanueva,
  Budzianowski, Mrk{\v{s}}i{\'c}, Wen, Ga{\v{s}}i{\'c}, and Young}]{pydial}
Stefan Ultes, Lina~M. Rojas-Barahona, Pei-Hao Su, David Vandyke, Dongho Kim,
  I{\~n}igo Casanueva, Pawe{\l} Budzianowski, Nikola Mrk{\v{s}}i{\'c},
  Tsung-Hsien Wen, Milica Ga{\v{s}}i{\'c}, and Steve Young. 2017.
\newblock \href {https://aclanthology.org/P17-4013} {{P}y{D}ial: A multi-domain
  statistical dialogue system toolkit}.
\newblock In \emph{Proceedings of {ACL} 2017, System Demonstrations}, pages
  73--78, Vancouver, Canada. Association for Computational Linguistics.

\bibitem[{Vahdati and Ramsin(2024)}]{DBLP:conf/modelsward/VahdatiR24}
Adel Vahdati and Raman Ramsin. 2024.
\newblock \href
  {https://www.scitepress.org/publishedPapers/2024/124337/pdf/index.html}
  {Model-driven methodology for developing chatbots based on microservice
  architecture}.
\newblock In \emph{Proceedings of the 12th International Conference on
  Model-Based Software and Systems Engineering, {MODELSWARD} 2024, Rome, Italy,
  February 21-23, 2024}, pages 247--254. {SCITEPRESS}.

\bibitem[{Vinyals and Le(2015)}]{vinyals2015neuralconversationalmodel}
Oriol Vinyals and Quoc Le. 2015.
\newblock \href {https://arxiv.org/abs/1506.05869} {A neural conversational
  model}.
\newblock \emph{Preprint}, arXiv:1506.05869.

\bibitem[{Walker et~al.(1997)Walker, Litman, Kamm, and
  Abella}]{walker-etal-1997-paradise}
Marilyn~A. Walker, Diane~J. Litman, Candace~A. Kamm, and Alicia Abella. 1997.
\newblock \href {https://doi.org/10.3115/976909.979652} {{PARADISE}: A
  framework for evaluating spoken dialogue agents}.
\newblock In \emph{35th Annual Meeting of the Association for Computational
  Linguistics and 8th Conference of the {E}uropean Chapter of the Association
  for Computational Linguistics}, pages 271--280, Madrid, Spain. Association
  for Computational Linguistics.

\bibitem[{Wallace(2009)}]{wallace2009anatomy}
Richard~S Wallace. 2009.
\newblock \href
  {https://link.springer.com/chapter/10.1007/978-1-4020-6710-5_13} {\emph{The
  anatomy of ALICE}}.
\newblock Springer.

\bibitem[{Wang(1998)}]{wang1998event}
Kuansan Wang. 1998.
\newblock \href {https://www.isca-archive.org/icslp_1998/wang98b_icslp.html}
  {An event driven model for dialogue systems.}
\newblock In \emph{ICSLP}.

\bibitem[{Washizaki(2024)}]{swebok4}
Hironori Washizaki, editor. 2024.
\newblock \href
  {https://www.computer.org/education/bodies-of-knowledge/software-engineering}
  {\emph{Guide to the Software Engineering Body of Knowledge v4.0}}.
\newblock IEEE Computer Society.

\bibitem[{Williams et~al.(2016)Williams, Raux, and
  Henderson}]{williams2016dialog}
Jason~D Williams, Antoine Raux, and Matthew Henderson. 2016.
\newblock \href {https://journals.uic.edu/ojs/index.php/dad/article/view/10729}
  {The dialog state tracking challenge series: A review}.
\newblock \emph{Dialogue \& Discourse}, 7(3):4--33.

\bibitem[{Xu et~al.(2022)Xu, Szlam, and Weston}]{xu-etal-2022-beyond}
Jing Xu, Arthur Szlam, and Jason Weston. 2022.
\newblock \href {https://doi.org/10.18653/v1/2022.acl-long.356} {Beyond
  goldfish memory: Long-term open-domain conversation}.
\newblock In \emph{Proceedings of the 60th Annual Meeting of the Association
  for Computational Linguistics (Volume 1: Long Papers)}, pages 5180--5197,
  Dublin, Ireland. Association for Computational Linguistics.

\bibitem[{Ye and Li(2020)}]{chatbot-security-privacy}
Winson Ye and Qun Li. 2020.
\newblock \href {https://doi.org/10.1109/SEC50012.2020.00057} {Chatbot security
  and privacy in the age of personal assistants}.
\newblock In \emph{2020 IEEE/ACM Symposium on Edge Computing (SEC)}, pages
  388--393.

\bibitem[{Yoshikawa et~al.(2024)Yoshikawa, Saeki, Takatsu, Kurata, and
  Matsuyama}]{yoshikawa:dialops}
Sadahiro Yoshikawa, Mao Saeki, Hiroaki Takatsu, Fuma Kurata, and Yoichi
  Matsuyama. 2024.
\newblock \href {https://doi.org/10.11517/jsaislud.102.0_235} {Dialops:
  Continuous development and operational management framework for large-scale
  dialogue systems}.
\newblock \emph{JSAI Technical Report, SIG-SLUD}, 102:235--240.
\newblock (in Japanese).

\bibitem[{Yoshino et~al.(2024)Yoshino, Chen, Crook, Kottur, Li, Hedayatnia,
  Moon, Fei, Li, Zhang, Feng, Zhou, Kim, Liu, Jin, Papangelis, Gopalakrishnan,
  Hakkani-Tur, Damavandi, Geramifard, Hori, Shah, Zhang, Li, Sedoc, D'Haro,
  Banchs, and Rudnicky}]{10174647}
Koichiro Yoshino, Yun-Nung Chen, Paul Crook, Satwik Kottur, Jinchao Li, Behnam
  Hedayatnia, Seungwhan Moon, Zhengcong Fei, Zekang Li, Jinchao Zhang, Yang
  Feng, Jie Zhou, Seokhwan Kim, Yang Liu, Di~Jin, Alexandros Papangelis,
  Karthik Gopalakrishnan, Dilek Hakkani-Tur, Babak Damavandi, Alborz
  Geramifard, Chiori Hori, Ankit Shah, Chen Zhang, Haizhou Li, João Sedoc,
  Luis~F. D'Haro, Rafael Banchs, and Alexander Rudnicky. 2024.
\newblock \href {https://doi.org/10.1109/TASLP.2023.3293030} {Overview of the
  tenth dialog system technology challenge: {DSTC10}}.
\newblock \emph{IEEE/ACM Transactions on Audio, Speech, and Language
  Processing}, 32:765--778.

\bibitem[{Zafar et~al.(2024)Zafar, Parthasarathy, Van, Shahid, Khan, and
  Shahid}]{bdcc8060070}
Ahtsham Zafar, Venkatesh~Balavadhani Parthasarathy, Chan~Le Van, Saad Shahid,
  Aafaq~Iqbal Khan, and Arsalan Shahid. 2024.
\newblock \href {https://doi.org/10.3390/bdcc8060070} {Building trust in
  conversational {AI}: {A} review and solution architecture using large
  language models and knowledge graphs}.
\newblock \emph{Big Data and Cognitive Computing}, 8(6).

\bibitem[{Zhao et~al.(2016)Zhao, Lee, and Eskenazi}]{zhao-etal-2016-dialport}
Tiancheng Zhao, Kyusong Lee, and Maxine Eskenazi. 2016.
\newblock \href {https://doi.org/10.18653/v1/W16-6007} {{D}ial{P}ort: A general
  framework for aggregating dialog systems}.
\newblock In \emph{Proceedings of the Workshop on Uphill Battles in Language
  Processing: Scaling Early Achievements to Robust Methods}, pages 32--34,
  Austin, TX. Association for Computational Linguistics.

\bibitem[{Zhu et~al.(2023)Zhu, Geishauser, Lin, van Niekerk, Peng, Zhang, Feng,
  Heck, Lubis, Wan, Zhu, Gao, Gasic, and Huang}]{zhu-etal-2023-convlab}
Qi~Zhu, Christian Geishauser, Hsien-chin Lin, Carel van Niekerk, Baolin Peng,
  Zheng Zhang, Shutong Feng, Michael Heck, Nurul Lubis, Dazhen Wan, Xiaochen
  Zhu, Jianfeng Gao, Milica Gasic, and Minlie Huang. 2023.
\newblock \href {https://doi.org/10.18653/v1/2023.emnlp-demo.9} {{C}onv{L}ab-3:
  A flexible dialogue system toolkit based on a unified data format}.
\newblock In \emph{Proceedings of the 2023 Conference on Empirical Methods in
  Natural Language Processing: System Demonstrations}, pages 106--123,
  Singapore. Association for Computational Linguistics.

\end{thebibliography}

\appendix

\section{Various Architectures of Dialogue Systems} \label{appendix:architecture}

This appendix introduces dialogue system architectures that could not be covered in detail in the main text.

\paragraph{Domain and language portability}

Commercial dialogue systems are often developed for specific domains and languages. Reducing the effort required to adapt such systems to other domains or languages is desirable. This property is referred to as domain portability and language portability. Many architectures have been proposed to support portability \cite{dahlback2003experiences, flycht-eriksson-jonsson-2000-dialogue, allen:nle00, LEE2009466, truong-etal-2017-maca}. These architectures typically separate domain- and language-independent processing from domain- and language-specific rules, enabling easier adaptation by replacing only the domain/language-dependent components.

\paragraph{Multi-domain dialogue system architecture}

In multi-domain dialogue systems, distributed architectures are commonly used, where each domain has its own dialogue management component \cite{ONEILL_McTEAR_2000, isobe:interspeech03, hartikainen04_interspeech, komatani-etal-2006-multi, NAKANO2011248}. Distributed architectures reduce system construction and maintenance costs by allowing easy addition and removal of domains.

\paragraph{Spoken and multimodal dialogue system architecture}

Unlike text-based systems, spoken and multimodal dialogue systems must handle irregular interactions, such as user barge-ins during system speech. To support such behavior, multi-layered architectures have been proposed \cite{raux:asru2007}. These architectures separate real-time processing (e.g., barge-in handling) in lower layers from symbolic-level dialogue management and response timing decisions in upper layers, thereby improving modularity and reducing coupling. The architecture of the SmartKom multimodal system \cite{Herzog2006} allows for parallel execution of input understanding and action generation, as well as incremental processing.

\paragraph{Client-server architecture}

As speech recognition and statistical language understanding have become more computationally intensive, offloading these components to a server has become more feasible in terms of speed and power consumption. Many commercial systems, such as Apple Siri, adopt this client-server architecture. A common issue with this architecture is response delay due to client-server communication. \citet{fuchs-etal-2012-scalable} propose an architecture in which speech recognition, language understanding, and dialogue management are executed on the server side, showing only minimal delays compared to desktop versions.

\paragraph{End-to-end systems and large language models}

Since the introduction of neural dialogue models \cite{lowe-etal-2015-ubuntu, vinyals2015neuralconversationalmodel}, many end-to-end dialogue systems have been proposed. These models allow systems to be trained solely on human dialogue data, resulting in simpler architectures from a software engineering perspective. However, achieving higher performance has led to increased architectural complexity. For example, BlenderBot 2\footnote{\url{https://parl.ai/projects/blenderbot2/}} introduces modules for long-term memory and web search \cite{komeili-etal-2022-internet, xu-etal-2022-beyond}. Recently, dialogue systems using a single prompt and an LLM have been studied \cite{madotto2021fewshotbotpromptbasedlearning}, while more advanced architectures use multiple modules that each invoke an LLM \cite{hudecek-dusek-2023-large}.
A system architecture that combines LLMs with knowledge graphs to improve response accuracy has also been proposed \cite{bdcc8060070}.
In addition, RAG (Retrieval-Augmented Generation) is widely used in LLM-based question-answering systems to leverage external knowledge. Various architectural variations of RAG have been proposed \cite{gao2024retrievalaugmentedgenerationlargelanguage}.

\paragraph{Spoken dialogue foundation models}

While LLMs are primarily text-based, foundation models tailored for spoken dialogue are also being explored. Moshi is a full-duplex speech dialogue model that enables natural speaker turn-taking within a single module \cite{defossez2024moshispeechtextfoundationmodel}.

\section{Dialogue System Development Tools} \label{appendix:tools}

This appendix lists additional dialogue system development tools and frameworks not fully covered in the main text.

\paragraph{Statistical dialogue systems}

To build dialogue systems based on statistical models such as those using reinforcement learning, model training is essential. Tools have been developed to facilitate such processes, including: PyDial \cite{pydial},
OpenDial \cite{lison-kennington-2016-opendial}, and
ConvLab \cite{zhu-etal-2023-convlab}.

\paragraph{Task-oriented text dialogue systems}

For task-oriented text-based systems, Trindikit \cite{LARSSON_TRAUM_2000} supports dialogue management based on the information state update approach. More recently, Rasa Open Source \cite{bocklisch2017rasaopensourcelanguage} has enabled the creation of diverse systems through interchangeable modules in a pipeline architecture.

For building task-oriented dialogue systems using LLMs, tools such as Conversation Routines \cite{robino2025conversationroutinespromptengineering},
InstructTODS \cite{chung2023instructtodslargelanguagemodels}, and
DLGNet-Task \cite{olabiyi2020dlgnettaskendtoendneuralnetwork}
have been proposed.

\paragraph{Open-domain chat-oriented dialogue systems}

For open-domain chat-oriented text-based dialogue systems, AIML interpreters \cite{wallace2009anatomy} offer a rule-based approach. Tools like ParlAI \cite{miller2018parlaidialogresearchsoftware} allow for the development of chit-chat systems using deep neural networks. Additionally, open-source tools such as Chatbot UI\footnote{\url{https://github.com/mckaywrigley/chatbot-ui}} support chat system development based on LLMs.

\paragraph{Spoken dialogue systems}

In addition to the VoiceXML interpreter mentioned in the main text, various tools have been developed for building spoken dialogue systems. 
SpeechBuilder \cite{DBLP:conf/interspeech/GlassW01} enables the construction of module-specific knowledge in the system using simple descriptions. 
ARIADNE \cite{denecke-2002-rapid} allows systems to be built without programming the dialogue manager by specifying access rules for backend databases, ontologies, and language understanding rules. 
Olympus \cite{bohus-etal-2007-olympus} features easy implementation of error handling.
Retico \cite{michael-2020-retico} is a tool for creating incremental spoken dialogue systems.

\paragraph{Multimodal dialogue systems}

The CSLU Toolkit \cite{cslu-tookit,sutton1998universal}, mentioned in the main text, enables the development of multimodal systems using visual agent representations. Other tools developed for agent visualization and multimodal interaction include Galatea Toolkit \cite{Kawamoto2004},
MMDAgent \cite{mmdagent}, and
VHToolkit \cite{hartholt:iva13,hartholt:iva22}.

Additional multimodal dialogue development tools include IrisTK \cite{skantze-johansson-2015-modelling} for multi-party human–robot interaction, and
\textbackslash psi \cite{bohus:arxiv21} for more flexible multimodal system development.
Remdis \cite{chiba:iwsds24} facilitates the development of dialogue systems that achieve incremental dialogue processing and the control of a virtual agent.
ScriptBoard \cite{lala-etal-2025-scriptboard} is a tool for building multi-party dialogue robots, providing a scenario editor.

\end{document}